\columnwidth\textwidth

\documentstyle[dina4,12pt]{article}

\begin{document}
\renewcommand{\theequation}{\arabic{section}.\arabic{equation}}
\parskip=4pt plus 1pt
\textheight=8.7in
\textwidth=6.0in
\newcommand{\be}{\begin{equation}}
\newcommand{\ee}{\end{equation}}
\newcommand{\bea}{\begin{eqnarray}}
\newcommand{\eea}{\end{eqnarray}}
\newcommand{\co}{\; \; ,}
\newcommand{\nn}{\nonumber \\}
\newcommand{\hlogm}{\ln \frac{M^2}{\mu^2}}
\newcommand{\apbc}{(\alpha + \beta)^C}
\newcommand{\ambc}{(\alpha - \beta)^C}
\newcommand{\apbn}{(\alpha + \beta)^N}
\newcommand{\ambn}{(\alpha - \beta)^N}
\newcommand{\ambcn}{(\alpha - \beta)^{C,N}}
\newcommand{\hmpp}{M_{\pi}}
\newcommand{\mppq}{M_{\pi}^2}
\newcommand{\p}[1]{(\ref{#1})}
\newcommand{\D}[1]{{\cal D}^{#1}}
\newcommand{\gag}{$\gamma \gamma \rightarrow \pi^0 \pi^0$}
\newcommand{\gpgpz}{$\gamma\pi^0 \rightarrow \gamma \pi^0$}
\newcommand{\gpgp}{$\gamma\pi \rightarrow \gamma \pi$}
\newcommand{\ggppz}{$\gamma\gamma\rightarrow\pi^0\pi^0$}
\newcommand{\ggpp}{$\gamma\gamma\rightarrow\pi\pi$}
\newcommand{\mscript}[1]{{\mbox{\scriptsize #1}}}
\newcommand{\mtiny}[1]{{\mbox{\tiny #1}}}
\newcommand{\MS}{\mtiny{MS}}
\newcommand{\GeV}{\mbox{GeV}}
\newcommand{\MeV}{\mbox{MeV}}
\newcommand{\keV}{\mbox{keV}}
\newcommand{\ren}{\mtiny{ren}}
\newcommand{\kin}{\mtiny{kin}}
\newcommand{\hint}{\mtiny{int}}
\newcommand{\tot}{\mtiny{tot}}
\newcommand{\CHPT}{\mtiny{CHPT}}
\newcommand{\DISP}{\mtiny{DISP}}
\newcommand{\CA}{\mtiny{CA}}
\newcommand{\scs}{\co \;}
\newcommand{\sem}{ \; \; ; \;}
\newcommand{\per}{ \; .}
\newcommand{\la}{\langle}
\newcommand{\ra}{\rangle}
\newcommand{\unith}{{\bf{\mbox{1}}}}
\begin{titlepage}

\def\mytoday#1{{ } \ifcase\month \or
 January\or February\or March\or April\or May\or June\or
 July\or August\or September\or October\or November\or December\fi
 \space \number\year}

\noindent
\hspace*{11cm} BUTP-93/18\\
\hspace*{11cm} LNF-93/077 (P)\\
\hspace*{11cm} PSI-PR-93-17\\
\vspace*{0.5cm}
\begin{center}
{{\Large{\bf{{Low-energy photon-photon collisions \\ to two-loop order}}
}}}$^\sharp$

\vspace{1cm}

S. Bellucci$^a$, J. Gasser$^b$ and M.E. Sainio$^{b,c,d}$

\vspace{1.2cm}

\mytoday \\

\vspace{.4cm}

\end{center}
\begin{abstract}

\noindent
We evaluate the amplitude for $\gamma \gamma \rightarrow \pi^0 \pi^0$ to two
loops in chiral perturbation theory. The three new  counterterms which enter
at this order in the low-energy expansion  are
estimated with  resonance saturation. We find that the
cross section agrees rather well with the available data and with dispersion
theoretic calculations even substantially above threshold.
Numerical results for the Compton cross section and for the neutral pion
polarizabilities are also given to two-loop accuracy.

\noindent
{\underline{\hspace{5cm}}}\\

\noindent
$\sharp$ Work supported in part by Schweizerischer Nationalfonds and by the EEC
Human Capital and Mobility Program.

\vspace{.15cm}

\noindent
a) INFN-Laboratori Nazionali di Frascati,
P.O.Box 13,
I-00044 Frascati, Italy.

\noindent
b) Institute for Theoretical Physics, University of Bern,
Sidlerstrasse 5, CH-3012 Bern, Switzerland.

\noindent
c) Paul Scherrer Institut,
CH-5232 Villigen PSI, Switzerland.

\noindent
d) During the academic year 1992-93 on leave of absence from
Dept. of Theo\-retical Physics, University of Helsinki, Finland.

\vspace{.15cm}
\noindent
e-mail: Bellucci@lnf.infn.it; Gasser@itp.unibe.ch; Sainio@finuhcb.bitnet
\end{abstract}
\end{titlepage}

\newpage

\setcounter{page}{2}
\setcounter{section}{0}
\setcounter{equation}{0}
\setcounter{subsection}{0}

         \section{Introduction\label{in}}

         The cross section for $\gamma \gamma \rightarrow \pi^0\pi^0$ and for
$\gamma \gamma \rightarrow \pi^+\pi^-$
has been
 calculated some time ago \cite{bico,dhlin} in the framework of chiral
perturbation
         theory (CHPT) \cite{wein79}-\cite{review} and of dispersion relations.
 For charged pion-pair       production, the chiral calculation \cite{bico}
at next-to-leading order
is in good agreement with
         the Mark II data \cite{dcharged} in the low-energy region.
 On the other hand, for
         $\gamma \gamma \rightarrow \pi^0\pi^0$, the one-loop
         prediction \cite{bico,dhlin} disagrees with the Crystal Ball data
\cite{cball}
 and with dispersion theoretic
         calculations \cite{goble}-\cite{kaloshin}
 even near
 threshold.

         In the process $\gamma \gamma \rightarrow \pi^+\pi^-$,
         the leading contribution\footnote{
         In this article, we denote the first nonvanishing
contribution to any quantity by  "the leading-order term", independently of
whether it starts out at tree level or at higher order in the chiral
expansion.}
 is generated by tree
         diagrams.
 One has a
         control on higher order corrections in this case, in the sense that
it is
         explicitly seen  that the one-loop graphs do not modify the
         tree amplitude very strongly near  threshold \cite{bico}.
 Tree diagrams are absent for $\gamma \gamma \rightarrow
         \pi^0 \pi^0$ which starts out with one-loop graphs.
 It is the aim of
this article to establish the region of validity of the
chiral representation of this process
by evaluating the amplitude at two-loop order.

         Is a next-to-leading order calculation
         sufficient in this case? If the corrections are large, the
         reliability of the result is certainly doubtful. However, a glance at
         the data shows that the corrections needed to bring CHPT and
         experiment into agreement are not large--a 25-30\% change in
         amplitude is  sufficient. Corrections of this size are
         rather normal in reactions where pions in an isospin zero
         $S$-wave state are present \cite{tr}. As an example we mention the
         isospin zero $S$-wave $\pi \pi$ scattering length, whose
         tree-level value \cite{weipp} receives a 25\% correction from one-loop
         graphs \cite{glan}. Corrections of a similar size are present
         in the scalar form factor of the pion \cite{gmff}.

         The amplitude for $\gamma \gamma \rightarrow \pi^0 \pi^0$ also
         describes Compton scattering on neutral pions
          by analyticity and crossing. Do sizeable corrections
         in $\gamma \gamma \rightarrow \pi^0 \pi^0$ then also show up
         in $\gamma \pi^0 \rightarrow \gamma \pi^0$ ?
 Since
there are no strongly interacting particles in the final state in this
         case, one might be led to suspect that the one-loop amplitude
         is a good approximation for this reaction. We find it
         interesting  that this is not the case--the
          corrections to the leading-order term are in fact very large in this
channel.

         The electromagnetic polarizabilities characterize  aspects
         of the inner structure of hadrons. With the two-loop
         expression for the amplitude at hand, it is straightforward
         to evaluate the polarizabilities  at
         next-to-leading order in the quark mass expansion.
         Renormalization group arguments show that this expansion
         contains logarithmic singularities of the type $M_\pi \ln^2
         M^2_\pi$ and $M_\pi \ln M_\pi$, and an order of magnitude
         estimate reveals that these  contributions may easily be as
         large as the leading-order term, unless the relevant
         Clebsch-Gordan coefficient is small. We find
         that the latter is the case.

Recently, a reformulation of CHPT has been given \cite{reffks}, where the
effective lagrangian includes into each order additional terms which in the
standard CHPT (considered here) are relegated to higher orders. To all orders,
the two perturbative schemes are identical--in each finite order, they may,
however, substantially differ. For an analysis of the process $\gamma \gamma
\rightarrow \pi^0 \pi^0$ in this generalized framework we refer the reader to
Ref. \cite{pipiks}.

         The article is organized as follows. In section 2, we set up
         the notation. In section 3 we describe
          the low-energy expansion in a general manner and outline the
specific procedure for the two-loop case in sections 4 and 5. The low-energy
constants which occur in the amplitude for $\gamma \gamma \rightarrow \pi^0
\pi^0$ at two-loop order are determined in
section 6. Section 7 contains
a discussion of the amplitude and of the cross section  at two-loop order.
The Compton amplitude and the pion polarizabilities are described in section 8,
whereas section 9 is devoted to a comparison of the chiral expansion with the
dispersive analysis of $\gamma \gamma \rightarrow \pi^0 \pi^0$ by
 Donoghue and Holstein \cite{dohod}.
Finally, a summary and
concluding remarks are presented in section 10.

\setcounter{equation}{0}
\setcounter{subsection}{0}

\section{Kinematics\label{nk}}

         The matrix element for pion production
         \be
         \gamma(q_1) \gamma(q_2) \rightarrow \pi^0(p_1) \pi^0(p_2)
         \ee
         is given by
         \be
         < \pi^0(p_1) \pi^0(p_2) \mbox{out} \mid \gamma(q_1)
         \gamma(q_2) \mbox{in} > = i(2 \pi)^4 \delta^4(P_f-P_i)
         T^N \co
         \ee
         with
         \bea
         T^N
         & = & e^2 \epsilon^\mu_1 \epsilon^\nu_2 V_{\mu \nu} \co
         \nn
         V_{\mu \nu} & = & i \int dxe^{-i(q_1x+q_2y)}
         < \pi^0(p_1) \pi^0(p_2) \mbox{out} \mid Tj_\mu(x) j_\nu(y)
         \mid 0 >.
         \eea
         Here $j_\mu$ is the electromagnetic current, and $\alpha =
         e^2/4 \pi \simeq 1/137$. The decomposition of the correlator
         $V_{\mu \nu}$ into Lorentz invariant amplitudes reads with
         $q^2_1 = q^2_2 = 0$ (see appendix \ref{ad})
         \bea
         V_{\mu \nu} &=& A(s,t,u) T_{1\mu \nu} + B(s,t,u) T_{2\mu \nu} +
          C(s,t,u) T_{3 \mu
         \nu} + D(s,t,u) T_{4\mu \nu} \co
         \nn
         T_{1\mu \nu}& =& \frac{s}{2} g_{\mu \nu} - q_{1\nu} q_{2\mu} \co
         \nn
         T_{2\mu \nu} &=&  2 s \Delta_\mu \Delta_\nu -\nu^2 g_{\mu \nu}
          - 2 \nu (q_{1\nu}
         \Delta_\mu
         - q_{2\mu} \Delta_\nu) \co
         \nn
         T_{3\mu \nu} &=& q_{1\mu} q_{2\nu} \co
         \nn
         T_{4\mu \nu}&=& s(q_{1\mu} \Delta_\nu - q_{2\nu} \Delta_\mu)
         - \nu(q_{1\mu} q_{1 \nu} + q_{2\mu} q_{2\nu}) \co
         \nn
         \Delta_\mu &=& (p_1 -p_2)_\mu \co
         \label{na1}
         \eea
         where
         \bea
         s&=& (q_1 + q_2)^2,\; t = (p_1 -q_1)^2, \; u = (p_2 - q_1)^2 \co
         \nn
         \nu &=& t-u \co
         \label{na2}
         \eea
         are the standard Mandelstam variables. The tensor $V_{\mu
         \nu}$  satisfies the Ward identities
         \be
         q^\mu_1 V_{\mu \nu} = q^\nu_2 V_{\mu \nu} = 0 \per
         \ee
         The amplitudes $A$ and $B$  are
         analytic functions of the variables $s,t$ and $u$, symmetric under
crossing $(t,u)\rightarrow (u,t)$. The quantities
 $C$ and $D$ do not contribute to the process considered here
(gauge invariance).

         It is useful to introduce in addition the helicity amplitudes
         \bea
         H_{++} &=& A + 2 (4M^2_\pi - s) B \co
         \nn
         H_{+-} &=&  \frac{8(M_\pi^4- tu)}{s} B \per
         \label{na3}
         \eea
         The helicity components $H_{++}$ and $H_{+-}$ correspond to
         photon helicity differences $\lambda = 0,2$, respectively.
         They have partial wave expansions involving even $J \geq
         \lambda$,
         \bea
         H_{++} &=& \sum_{J=0,2,4\ldots} h_+^J(s) d_{00}^J (\cos \theta)
          \co
         \nn
         H_{+-} &=&   \sum_{J=2,4,6\ldots} h_-^J (s) d_{20}^J
         (\cos \theta) \co
\label{na4}
 \eea
         where $\theta$ is the scattering angle in the center-of-mass system,
$\vec{q}_1\cdot \vec{p}_1=|\vec{q}_1||\vec{p}_1|\cos \theta$.

         With our normalization of states $< \vec{p}_1 \mid \vec{p}_2> =
         2(2 \pi)^3 p_1^0 \delta^3 (\vec{p}_1 - \vec{p}_2)$, the
         differential cross section for unpolarized photons in the
         center-of-mass system is
         \bea
         {\frac{d \sigma}{d \Omega}}^
{\gamma \gamma \rightarrow \pi^0 \pi^0} &=& \frac{\alpha^2s}{64} \beta(s)
         H(s,t)\co
         \nn
         H(s,t) &=& \mid H_{++} \mid^2 + \mid H_{+-} \mid^2 \co \nn
         \beta(s)& =&
         (1-4 M^2_\pi/s)^{1/2}.
         \eea
         The amplitude for Compton scattering
         $$
         \gamma (q_1) \pi^0(p_1) \rightarrow \gamma(q_2) \pi^0(p_2)
         $$
 may be obtained by
         crossing. In the center-of-mass system, the cross section for
unpolarized photons is
         \be
         {\frac{d \sigma}{d \Omega}}^{\gamma \pi^0 \rightarrow \gamma \pi^0}
         = \frac{\alpha^2}{16 \bar{s}} \bar{t}^2 H (\bar{t}, \bar{s})\co
         \ee
         with
         $$
         \bar{s} = (q_1 + p_1)^2 \scs \bar{t} = (q_2 - q_1)^2 \per
         $$
 Finally, the optical theorem in the Compton channel reads with our phase
convention
\be
e^2{\mbox{Im}} B|_{s=0,t=\bar{s}} =
\frac{1}{4(\bar{s}-M_\pi^2)}\sigma_\tot^{\gamma \pi^0}
(\bar{s}) \;.
\label{na10}
\ee
This relation fixes the  phase of $A$ through Eq. (\ref{na1}).

  The physical region for the reactions $\gamma \gamma \rightarrow \pi^0 \pi^0$
 and $\gamma \pi^0 \rightarrow \gamma \pi^0$
 is displayed in Fig. 1,
       where we also indicate with  shaded lines the nearest
         singularities in the amplitudes $A$
         and $B$. These singularities are generated by two-pion
         intermediate states in the $s,t$ and $u$ channel.

\setcounter{equation}{0}
\setcounter{subsection}{0}

         \section{Low-energy expansion\label{le}}
          We consider QCD with two flavours in
         the isospin symmetry limit $m_u = m_d = \hat{m}$ and equip
         the underlying lagrangian with hermitean, colour neutral
         external fields $v,a,s$ and $p$ in the standard manner,
         \be
         {\cal L} = {\cal L}^0_{\rm{QCD}} + \bar{q}
         \gamma^\mu
         (v_\mu + a_\mu \gamma_5) q - \bar{q} (s-i \gamma_5 p) q \per
          \label{le0}
\ee
         Here ${\cal L}^0_{\mbox{\tiny{QCD}}}$ denotes the QCD
         lagrangian at zero quark mass, whereas $\hat{m}$ is contained in the
         scalar field $s(x)$. The lagrangian (\ref{le0}) is invariant under
         local $SU(2)_L\times SU(2)_R\times U(1)$
         transformations
         \be
         q \rightarrow \frac{1}{2} [(1+ \gamma_5) g_R + (1-\gamma_5)
         g_L] q
         \ee
         with
         \bea
          g_{R,L}&=&e^{i\phi}V_{R,L} \co \nn
  V_{R,L} &\in& \mbox{SU}(2)\co \nn
         \phi &=& \mbox{diag}(\phi_0, \phi_0) \co \phi_0 \in {\bf{{\mbox{R}}}}
         \co
    \eea
         provided that the external fields are subject to the gauge
         transformation
         \bea
         r'_\mu &=& g_R r_\mu g^\dagger_R + i g_R \partial_\mu g^\dagger_R
         \co
         \nn
         l'_\mu &=& g_L l_\mu g^\dagger_L + ig_L \partial_\mu g^\dagger_L
         \co
         \nn
         s'+ip' &=& g_R (s+ ip) g_L^\dagger \co
         \nn
         r_\mu &=& v_\mu + a_\mu \co \; l_\mu = v_\mu - a_\mu \per
         \label{le1}
         \eea
         Since the charge is not a generator of $SU(2)$,  we consider in the
following the case
         \be
         \la a_\mu \ra = 0 \co \; \la v_\mu \ra \neq 0 \co
         \label{le2}
         \ee
         where $\la A\ra$ denotes the trace of the matrix $A$. The
         condition (\ref{le2}) is consistent with the transformation
         law (\ref{le1}). The Green functions of the theory are
         generated by the vacuum-to-vacuum amplitude
         \be
         e^{iZ(v,a,s,p)} = < 0_{\mbox{\tiny{out}}} \mid
         0_{\mbox{\tiny{in}}} >_{v,a,s,p}.
         \ee
         The generating functional $Z$ admits an expansion in powers
         of the external momenta and of the quark
 masses \cite{wein79}-\cite{glnp},
         \be
         Z = Z_2 + Z_4 + Z_6 + \ldots \co
         \label{le3}
         \ee
         where $Z_{n}$ denotes a term of order $E^{n}$. We write
         the corresponding expansion of the amplitudes as
         \be
         I = I_2 + I_4 +I_6 + \ldots \sem I = V^{\mu \nu}, A, B \co
         \label{le4}
         \ee
         where it is understood that $Z_n$ generates $I_n$\footnote{Notice
 that $I_n$
is not of order $E^n$.}. To
 calculate $V^{\mu \nu}$, we set
         \bea
         s =  \hat{m}\unith \scs
         v_\mu = Q \bar{v}_\mu \scs
         p = \tau^3 \bar{p} \scs
         a_\mu = 0 \co
         \label{le5}
         \eea
         where
        \be
         Q = \frac{1}{3} \mbox{diag} (2,-1)
         \ee
         is the charge matrix, and
          where $\bar{v}_\mu$ and $\bar{p}$ denote flavour neutral
         external fields. $V^{\mu \nu}$ is obtained from the term of
         order $\bar{v}^2 \bar{p}^2$ in $Z$.

         \subsection{Terms at order $E^2$}

         In the meson sector, $Z_2$ is given by the classical action
         \be
         Z_2 = \int d x {\cal L}_2 (U, v,a,s,p) \co
         \label{le6}
         \ee
         where ${\cal L}_2$ is the nonlinear $\sigma$-model lagrangian
         \be
         {\cal L}_2 = \frac{F^2}{4} \la D_\mu UD^\mu U^\dagger + \chi^\dagger U
 + \chi U^\dagger\ra,
         \label{le7}
         \ee
         evaluated at the solution to the classical equation of motion
$\delta\int {\cal L}_2 =0$.
         The $2\times2$ unitary matrix $U$ contains the pion fields,
         \bea
         U &=& \sigma + i \frac{\phi}{F} \scs \sigma^2 +
\frac{\phi^2}{F^2} = \unith \co
         \nn
         \phi &=& \left( \mbox{$\begin{array}{cc} \pi^0 & \sqrt{2} \pi^+
         \\ \sqrt{2} \pi^- &- \pi^0 \end{array}$} \right) = \phi^i \tau^i \; .
         \label{le8}
         \eea
         It transforms as
         \be
         U \stackrel{G}{\rightarrow}  g_R Ug_L^\dagger
         \ee
         under $G =SU(2)_L \times SU(2)_R\times U(1)$. The covariant
         derivative is
         \be
         D_\mu U = \partial_\mu U- i r_\mu U + i Ul_\mu \co
         \ee
         and the field $\chi$ denotes the combination

         \be
         \chi = 2 B(s+ i p).
         \ee
         $F$ is the pion decay constant in the chiral limit, $F_\pi = F
         (1 + O(\hat{m})), F_\pi \simeq 93$ MeV, and $B$ is related to
         the order parameter $< 0 \mid \bar{q} q \mid 0>$.
          The physical pion mass is
         \bea
         M_\pi^2 &=& M^2 (1 + O (\hat{m})) \co
         \nn
         M^2 &=& 2 \hat{m} B.
         \label{le9}
         \eea
         $  {\cal L}_2$ is referred to as the effective
         lagrangian at order $E^2$.

         The term of order $O(\bar{v}^2\bar{p}^2)$ in the classical action
$Z_2$ vanishes and, therefore,  one has
         \be
         A_2 = B_2 = V^{\mu \nu}_2 = 0 \per
         \ee

         \subsection{Higher orders in the energy expansion}

         At higher orders in the energy expansion, the effective
         lagrangian consists of a string of terms. Reintroducing momentarily
$\hbar$, one has
         \be
         {\cal L}_{\mbox{\tiny{eff}}} = {\cal L}_2 + \hbar {\cal
         L}_4 + \hbar^2 {\cal L}_6 + \cdots .
         \ee
         Here ${\cal L}_4$ contains all possible contributions with
         four derivatives, or two derivatives and one field $\chi$, or
         $\chi^2$, and similarly for the higher order terms (${\cal
         L}_4$ contains in addition the Wess-Zumino-Witten lagrangian
         ${\cal L}_{\mbox{\tiny{WZW}}}$ \cite{wessz}).
         The
         generating functional is given by
         \be
         e^{\frac{iZ}{\hbar} (v,a,s,p)} = \int [dU]
         e^{ {\frac{i}{\hbar} \int {\cal L}_{\mbox{\tiny{eff}}}} dx }\co
         \ee
         and its low-energy expansion is obtained from $Z = Z_2 +
         \hbar Z_4 + \cdots$ . One expands
         \be
         {\cal L}_I = \bar{{\cal L}}_I + C_I \xi + \frac{1}{2} \xi D_I
         \xi + E_I \xi^3 + F_I\xi^4 +\cdots \sem I = 2,4, \ldots \co
         \label{le10}
         \ee
         where $\bar{{\cal L}}_I$ denotes the lagrangian ${\cal L}_I$,
evaluated at the solution to the
         classical equation of motion $\delta \int {\cal{L}}_2=0$. (To
simplify the notation, we have dropped the $SU(2)$ - indices in $\xi$ and
in the operators $C_I,D_I,\ldots$ .)
 The fluctuation $\xi$ is of order
         $\hbar^{1/2}$.
 Then one obtains
         \bea
         e^{\frac{iZ}{\hbar}} &=& e^{\frac{i}{\hbar}
         S_{\mbox{\tiny{cl}}}}
         \int [d \xi] e^{\frac{i}{\hbar} \int \frac{1}{2} \xi D_2 \xi dx}
\triangle \co           \nn
        \triangle &=&1 - \frac{1}{2 \hbar^2}
         \int \left [(E_2 \xi^3 +\hbar C_4\xi)_x
(E_2 \xi^3 +\hbar C_4\xi)_y
\right] dxdy  \nn
         &&+\frac{i}{2\hbar}\int \left [2F_2\xi^4 + \hbar
            \xi D_4\xi \right]_x dx   +O(\hbar^2) \co
         \label{le11}
         \eea
         with $S_{\mbox{\tiny{cl}}} = \int dx \bar{{\cal
         L}}_{\mbox{\tiny{eff}}}$.

 At order $E^4$, this result amounts to
         evaluating one-loop graphs generated by ${\cal L}_2$ and adding
         the tree graphs from ${\cal L}_2 + \hbar{\cal L}_4$ \cite{glan}.
These contributions then add up to $Z_4$, which contains the leading-order
term  $V_4^{\mu \nu}$. It is a specific feature of the
process $\gamma \gamma \rightarrow \pi^0 \pi^0$ that the counterterms contained
in
${\cal L}_4$ do not contribute to $V_4^{\mu \nu}$--the sum of
the one-loop graphs is therefore ultraviolet finite \cite{bico,dhlin}.

The diagrams which generate $Z_6$   are
displayed in
Fig. 2. The solid-dashed lines stand for the propagator $D_2^{-1}$,
 and the framed symbols $I$ denote vertices
from ${\cal L}_I$ according to Eq. (\ref{le10}).

In order not to interrupt the argument, we relegate the discussion of the
leading contribution $V_4^{\mu \nu}$ to appendix \ref{ao} and continue in the
following
section with the evaluation of the next-to-leading order term $V_6^{\mu \nu}$.

         \section{Renormalization\label{tlr}}

         The evaluation of $V_6^{\mu \nu}$ is  complex. We outline
         in this and  in the following two sections the procedure -- omitting,
however,
all details.

         \subsection{The lagrangians ${\cal L}_4$ and ${\cal L}_6$}

         The lagrangian ${\cal L}_4$ contributes to $V^{\mu
         \nu}_6$
         through  one-loop diagrams, see Fig. 2. Its general form is
\cite{glan}
         \bea
         {\cal L}_4 &=& {\cal L}^{(4)} + {\cal L}_{\mbox{\tiny{WZW}}}
         \co
         \nn
         {\cal L}^{(4)} &=& \sum^{7}_{i=1} l_iP_i + \cdots \co
         \label{tl1}
         \eea
         where
         \bea
         P_1 &=& \frac{1}{4} \la u^\mu u_\mu \ra^2 \co \nn
         P_2 &=& \frac{1}{4} \la u_\mu
         u_\nu\ra \la u^\mu u^\nu \ra \co
         \nn
         P_3 &=& \frac{1}{16} \la \chi_+\ra^2 \co \nn
         P_4 &=& \frac{i}{4} \la u_\mu
         \chi^\mu _- \ra \co \nn
         P_5 &=& - \frac{1}{2} \la f_-^{\mu \nu} f_{- \mu \nu} \ra \co \nn
         P_6 &=&
         \frac{i}{4} \la f^{\mu \nu}_+ [u_\mu, u_\nu]\ra \co
         \nn
         P_7 &=& -\frac{1}{16}\la \chi_- \ra^2.
         \label{tl2}
         \eea
         Here we used the notation
         \bea
         u_\mu &=& i u^\dagger D_\mu Uu^\dagger = -i u D_\mu U^\dagger u =
u_\mu^\dagger \co
         \nn
         \chi_\pm &=& u^\dagger \chi u^\dagger \pm u \chi^\dagger u \co \nn
         \chi_-^\mu&=& u^\dagger D^\mu\chi u^\dagger -uD^\mu\chi^\dagger u \co
\nn
         f^{\mu \nu}_\pm &=& u F^{\mu \nu}_L u^\dagger \pm u^\dagger F^{\mu
\nu}_R
         u \co
         \label{tl3}
         \eea
         with $u^2=U$. The quantity $F^{\mu \nu}_R \; (F_L^{\mu \nu})$ stands
 for the field strength associated with the nonabelian external field
         $v_\mu + a_\mu \; (v_\mu - a_\mu)$.

The ellipsis in (\ref{tl1}) denotes polynomials in the external fields
which are independent of the pion variables. These do not contribute to
$S$-matrix elements and are therefore not needed in the following. Finally, the
anomaly term ${\cal{L}}_\mtiny{WZW}$ contributes
 to $V_8^{\mu \nu}$ \cite{wzbijn}. This is beyond the accuracy of
the low-energy expansion considered here.

The realization of $G$ on $u$ is
         \be
         u \stackrel{G}{\rightarrow} g_R u h^\dagger = hu g^\dagger_L
         \co
         \ee
         such that
         \be
         I \stackrel{G}{\rightarrow} hIh^\dagger
         \ee
         for the quantities in (\ref{tl3}). The low-energy constants $l_i$ are
          divergent, except $l_7$. They remove the ultraviolet
         divergences generated by the one-loop graphs--we discuss
         them in more detail below.

         In the construction of ${\cal L}^{(4)}$, the
         equation of motion $\delta \int {\cal{L}}_2=0$ has been
         used. It can be shown that adding terms to ${\cal L}^{(4)}$
         which vanish upon use of the equation of motion
         affects
 the generating functional at order $E^6$ by a local
         term\footnote{We thank G. Ecker for an explicit proof of this
 statement and for illuminating discussions concerning the material in section
4.2.}--these contributions may thus be omitted.

         The lagrangian ${\cal L}_6$ contributes a polynomial
         part to $V_6^{\mu \nu}$ which cancels the ultraviolet
         singularities generated by the two-loop diagrams. The general
         structure of ${\cal L}_6$ is not yet available in the
         literature \cite{bijn2l}.
          Concerning the present calculation, we note
         that the lagrangian
         \bea
         {\cal L}_6 &=& \frac{1}{F^2} \la f_{+\mu\rho} f_+^{\mu \sigma} +
           f_{-\mu\rho} f_-^{\mu \sigma} \ra T_\sigma^\rho
          + \cdots \co
         \nn
         T_{\rho\sigma} &=&  d_1\la u_\rho u_\sigma\ra + g_{\rho\sigma}
  \{d_2 \la u^\mu u_\mu \ra+ d_3 \la \chi_+\ra\}
         \eea
         generates a polynomial in $A_6, B_6$ which has the same
         structure as the divergent part in the two-loop contribution,
         \bea
         A_6 &=& \frac{20}{9F^4} [ 16 (d_3 -d_2) M^2 + (d_1 +
         8 d_2)s] + \cdots \co
         \nn
         B_6&=& -\frac{10}{9F^4} d_1 + \cdots.
         \eea
         We may therefore remove the divergences in $V_6^{\mu \nu}$ by
simply dropping the singular parts in  $A_6$ and $B_6$, see below.

   \subsection{Regularization and renormalization}

         We use dimensional regularization and set
         \be
         \omega = d-4 \co
         \ee
          where $d$ is the dimension of space-time.
         We introduce the renormalization scale $\mu$ such that the
         calculation is scale-independent at each step.
         In the following we outline the procedure \cite{renor}.

Consider the couplings $l_i$ in ${\cal L}^{(4)}$ which carry
         dimension (mass)$^{\omega}$. We treat $l_i$ as $\mu$-independent
         parameters by writing in the minimal subtraction scheme
         \be
         l_i = \mu^{\omega} \left \{ \frac{\delta_i}{ \omega} +
         l_i^{\MS} +  \omega l^{\MS}_{i1} + O(\omega^2) \right \} \; ; \;
         i = 1, \ldots, 7 \co
         \ee
         with
         \be
         \mu \frac{dl_i^{\MS}}{d \mu} = - \delta_i \co \;  \mu
         \frac{dl_{i1}^{\MS}}{d \mu} = - l_i^{\MS}\; .
         \ee
         The divergent terms $\delta_i / \omega$ remove
the one-loop
      singularities in $Z_4$ (the $\delta_i$ are related to the
$\gamma_i$ used in \cite{glan} by  $\delta_i = \gamma_i/16 \pi^2$).
 Since the $l_i$ occur
         in $Z_6$ via loop insertions, the constants $l_{i1}^{\MS}$ in
         general also contribute at two-loop order. For an illustration of the
renormalization procedure,
    we  consider the amplitude $F^4B$
         which has dimension (mass)$^{2\omega}$. We
         write for the contribution from the loops
\be
B=\frac{\mu^{2\omega}}{F^4}\left\{ {\mbox{Laurent-series of}}
\; (\mu^{-2\omega}F^4B) \;{\mbox{at}}\;\;\; \omega=0\right\} \per
\label{rr1}
\ee
 From ${\cal{L}}^{(4)}$, only $l_2$ contributes,
\be
B=\frac{1}{F^4}\left\{ M^{2\omega}f(s/M^2,t/M^2;\omega)
+M^{\omega}l_2 \; g(s/M^2,t/M^2;\omega)  \right\} \; ,
\label{rr2}
\ee
where
 $f$ and $g$ are singular as $\omega \rightarrow
0$,
\bea
f&=&\frac{f_{2}}{\omega^2} + \frac{f_{1}}{\omega} + f_0 +O(\omega) \co
\nn g&=& \frac{g_{1}}{\omega} + g_0 +\omega g_{+1}  +O(\omega^2) \; .
\label{tl4}
\eea
 The Laurent-series (\ref{rr1}) becomes
\be
B=\frac{\mu^{2\omega}}{F^4}\left\{ \frac{\beta_2}{\omega^2} +
\frac{\beta_{1}l_2^{\MS} +\beta_{1,1}} { \omega} +\beta^{\MS} +O(\omega)
\right\}\per
\label{tl5}
\ee
The residues of the pole terms are
\be
\beta_2=\frac{\delta_2 g_1}{2}\scs
\beta_{1} = g_{1} \scs
\beta_{1,1} = \delta_2 g_0 +f_{1} \scs
\label{tl6}
\ee
and for the finite part we find
\be
\beta^{\MS}=\frac{g_1}{8}\left\{-\delta_2 \hlogm +4
l_2^{\MS}\right\} \hlogm +c_1 \hlogm +c_2 \quad .
\label{tl61}
\ee
Here the $c_i$ stand for linear combinations of $f_1,\ldots,g_{+1}$.

Without having evaluated any Feynman diagram, we have already obtained
significant information on the structure of the two-loop result \cite{wein79}:
\begin{enumerate}
\item The residues of the pole-terms in (\ref{tl5}) are
polynomials in the external momenta and in the masses on general grounds
\cite{polynoms}. For dimensional reasons, these polynomials reduce to pure
numbers in the present case. In addition, from the cancellation
of the logarithmic terms
 which are generated by expanding  the factors $M^{2\omega}$
and $M^\omega$ in (\ref{rr2}), it follows that the residue of the double pole
in $f$ is proportional to the residue of the single pole in $g$,
 \be
 2f_2+\delta_2 g_1 = 0 \co
\label{tl6a0}
 \ee
 or
 \be 2\beta_2=\delta_2\beta_1 \; .
 \label{tl6a}
\ee

\item The amplitude $f$ contains a nonlocal singularity $f_1/\omega$ which is
generated by divergent subgraphs. This nonlocality  must be cancelled by the
nonlocal singular part $\delta_2 g_0/ \omega$ in the graphs generated by loops
with ${\cal{L}}^{(4)}$, in such a way that $\beta_{1,1} = \delta_2 g_0 +f_1$
becomes a pure number.

\item As is seen from Eq. (\ref{tl61}), the finite part
$\beta^{\MS}$ contains chiral logarithms $\ln^2M^2$ and $\ln M^2$. The
coefficient of the leading term $\ln^2M^2$ is proportional to the residue of
the single pole in $g$, whereas the  linear piece $\ln M^2$ is multiplied with
a nonlocal function. These singular terms cancel the chiral logarithms
in $c_2$ and thus generate  a smooth behaviour of the amplitude in the chiral
limit $\hat{m}\rightarrow 0$ (at fixed $s,t \neq 0$).

\item At the Compton
threshold $s=0,t=M^2$, the quantities $c_1$ and $c_2$ are independent of the
quark mass. Therefore, the chiral logarithms $\ln ^2M^2$ and $\ln M^2$ in
 Eq. (\ref{tl61}) remain, and   we conclude that the
 finite part $\beta^{\MS}|_{s=0,t=M^2}$ blows
up in the chiral limit. In other words, the slope of the form factor  $V_9=2sB$
(see appendix A) is infrared singular,
\be
 F^4\frac{dV_9}{ds}|_{s=0,t=M^2} =
\frac{g_1}{4}\left\{-\delta_2 \hlogm +4 l_2^{\MS}\right\}
\hlogm +\bar{c}_1\hlogm +\bar{c}_2
\quad ,
\ee
where the $\bar{c}_i$ are independent of the quark mass. Notice that $\ln^2M^2$
occurs together with $l_2^{\MS}\ln M^2$ in a particular combination which
is dictated by
Eq. (\ref{tl6a0}).
\end{enumerate}

\noindent
We now define the renormalized amplitude $B^\ren$ as
\be
B^\ren=\frac{1}{F^4}(\beta^{\MS} +b^{\MS} )\co
\ee
where the scale-dependence of $b^{\MS}$ is chosen
such that $B^\ren$ is
independent of $\mu$,
\be
\mu\frac{d}{d\mu}B^\ren=0\; .
\ee
(The low-energy parameter $b^\mtiny{MS}$ is the sum of the
finite pieces of the relevant counterterms at order $E^6$ in the effective
action.)

We have formulated the renormalization procedure in the minimal subtraction
scheme, where
$\ln(4\pi)$ and $\Gamma^{'}(1)$ occur. One may eliminate these terms in the
standard manner \cite{bardeen}. Below we use the conventions of
\cite{glan}.
The final result for $B_6$ contains one unknown new parameter
$b^r$,
\be
B_6=\frac{b^r}{(16\pi^2F^2)^2} + \cdots \; .
\label{tl7}
\ee
Analogously, the renormalized amplitude $A_6$ contains two unknown new
parameters
$a_1^r$ and $a_2^r$,
\be
A_6=\frac{a_1^r M^2 + a_2^r s}{(16\pi^2F^2)^2} +  \cdots \;.
\label{tl8}
\ee
The ellipses in (\ref{tl7}) and (\ref{tl8}) stand for the finite
contributions from the loop-integrals.

\setcounter{equation}{0}
\setcounter{subsection}{0}

         \section{Evaluation of diagrams\label{ed}}

         Here we discuss  further aspects of the two-loop
         calculation.

         \subsection{The diagrams}

         It is straightforward to generate from
         figure 2 the diagrams for $\gamma \gamma \rightarrow
\pi^0 \pi^0$ at two-loop order--one has simply to insert photon and pion
vertices in all possible ways. For illustration, we display one class of
graphs in Fig. 3. The solid (dashed) line denotes  charged
(neutral) pions. The four-point function on the right-hand side is the
$d$-dimensional elastic $\pi \pi$ scattering amplitude at one-loop accuracy,
 with two pions off-shell (the one-loop graphs for
 $\gamma \gamma \rightarrow
\pi^0 \pi^0$
are thus also included in Fig. 3).
        The symbol
         $d^Dl$ stands for integration over internal momenta with
         weight
         \be
         \frac{1}{[M^2_\pi - (l + q_1)^2] [M^2_\pi - (q_2 - l)^2 ]} \co
\label{fa0}
       \ee
         where $M^2_\pi$ denotes the physical pion mass in one-loop
         approximation,
         \bea
         M^2_\pi &=& M^2 \left[ 1 + \frac{M^2}{F^2}(2l_3^r
+\frac{1}{32\pi^2}\ln \frac{M^2}{\mu^2})
         + O(M^4) \right]  \; .
\label{fa1}
  \eea
The momenta of the charged pions  running in the
         loop are $(l+q_1)^\mu$ and $(q_2 - l)^\mu$. We do not
 display the remaining diagrams.

         \subsection{Numerical evaluation of diagrams}

         The derivative nature of the interaction makes the algebraic
         part of the calculation tedious. As for the numerical part,
         we have to evaluate the amplitudes in the physical region for
         $\gamma \gamma \rightarrow \pi^0 \pi^0$ where branch-points
         and cuts appear. These render the numerical evaluation of the
         Feynman  integrals nontrivial. To illustrate our procedure
to cope with this difficulty, we consider
          the two-loop box diagram Fig. 4a. We write
 for the subdiagram (enclosed by dash-dotted
         lines) a $d$-dimensional spectral representation
         \be
         J(\bar{t},d) = \int_{4M_\pi^2}^\infty \frac{d \sigma
\rho(\sigma,d)}{\sigma -
         \bar{t}}
\label{fa1a}
         \ee
         where $\bar{t}$ contains loop-momenta. This
         leads to a
spectral representation for the full diagram,
          \be
         \int^{\infty}_{4M_\pi^2} d \sigma \rho(\sigma,d) b(s,t,
         \sigma,d),
         \ee
         where $b(s,t, \sigma,d)$ denotes a one-loop $d$-dimensional box
diagram
(one of the internal lines carries mass $\sqrt{\sigma}$). After
removing the subdivergence generated by $J(\bar{t},d)$, we obtain
the finite part
 by writing a fixed-$t$ dispersion relation.
       This procedure allows one
to evaluate numerically the amplitude also in the region $s > 4M_\pi^2,t<
9M_\pi^2$.

         \subsection{Checks on the calculation}

         As is shown in appendix \ref{ad}, the most general expression for
the amplitude
         $\epsilon_1^\mu \epsilon_2^\nu V_{\mu \nu}$ contains five form
         factors which are linearly related
         through two Ward identities and through Bose symmetry,
\be
\begin{array}{rrrr}
         L_1(V_i) &\doteq& 2V_0 +sV_4 - (t-u) V_5 = 0 &\co \\
         L_2(V_i) &\doteq& sV_7 - (t-u) V_9 = 0 &\co   \\
         L_3(V_i) &\doteq& V_5+V_7 = 0 & .
 \end{array}
        \label{fa3}
         \ee
The amplitudes $A$ and $B$ may be obtained from $V_4$ and $V_9$,
\be
A=-V_4 \scs B=V_9/2s \; .
\ee
To get an optimal control of the calculation, we have fully evaluated all five
 form
factors in $d$ dimensions and have made the following
consistency checks of the results:
\begin{enumerate}
\item
We have verified that the relations  (\ref{fa3}) are satisfied numerically
within
machine accuracy in low dimensions  (below threshold).
\item
We have determined the divergence structure at $\omega \rightarrow 0$,
\be
V_i=\mu^{2\omega}\left\{\frac{P_i^{(2)}}{\omega^2} +\frac{P_i^{(1)}}
{\omega} +R_i
+O(\omega)\right\}\co
\ee
 and have
verified that the residues $P_i^{(k)}$ are  polynomials in
 the external
 momenta and in the masses. These polynomials obey Eq. (\ref{fa3}) as well,
\be
L_m(P_i^{(k)})=0 \; \; ; \; m=1,2,3 \; \; ; \; k=1,2\co
\ee
and  are related in the manner discussed in the previous section, see Eq.
 (\ref{tl6a}) for $P_9^{(1)}$ and $P_9^{(2)}$.
\item
We have checked that also the finite parts $R_i$ satisfy numerically
\be
L_m(R_i)=0 \sem m=1,2,3
\ee
within machine accuracy below threshold.

\noindent
At this stage, we have written fixed-$t$ dispersion relations for some of the
finite parts in the manner mentioned above. This allows one to evaluate
the complete amplitude in the physical region for pion-pair production.

\item
We have then worked out the $S$-wave projection $h_+^0(s)$   (\ref{na4}) and
have verified numerically that this amplitude has the correct phase at
$s>4M_\pi^2$, given by the (tree plus one-loop) elastic $\pi \pi$ scattering
$S$-wave phase shift (in the appropriate isospin decomposition).

\end{enumerate}
\noindent
In addition, we have made many other cross-checks.
\section {Low-energy constants at order $E^6$}

         Once the program described above is carried through, one ends
         up with ultraviolet finite and scale-independent amplitudes
         $A$ and $B$ which contain the parameters $F, M_\pi;\;
         l^r_i, l^r_{i1} \; (i=1,2,3,5,6);\; a_1^r, a_2^r$ and $b^r$.  $F$
is related to the physical pion decay constant $F_\pi$ \cite{glan},
\be
         F_\pi = F \left[ 1 + \frac{M^2}{ F^2} (l_4^r -\frac{1}{16\pi^2}\ln
\frac{M^2}{\mu^2}) +
      O(M^4)
         \right] \; \; .
\ee
We may therefore replace $F$ by $F_\pi$ at the expense of introducing $l_4^r$.
 The
expressions for the loop-amplitudes simplify if one uses \cite{glan} the
scale-independent parameters $\bar{l}_i$,
\be
l_i^r = \frac{\delta_i}{2}(\bar{l}_i +\ln \frac{M^2}{\mu^2})\; .
\label{lec1}
\ee
Their values are  displayed in column 2 of table 1. We note  that
$\bar{l}_5$ and $\bar{l}_6$
in the present application always  appear in the
combination
\be
\bar{l}_\triangle = \bar{l}_6 - \bar{l}_5 = 2.7\per
\label{rr10}
\ee
[The constant $\bar{l}_\triangle$ is related to the
low-energy couplings $L^r_9$ and $L^r_{10}$ which occur
in the $SU(3)_L \times SU(3)_R$ version of the one-loop amplitude $\gamma
\gamma
\rightarrow \pi^+\pi^-$ \cite{bico}
by $\bar{l}_\triangle =
 192 \pi^2 (L^r_9 + L^r_{10})$.]
Next we observe that we may
         absorb $l^r_{i1}$'s into the low-energy constants at order $E^6$,
because
         they contribute a polynomial piece only. We are therefore
         left with  $a_1^r,a_2^r$ and $b^r$ as the only new unknowns.
 We estimate these in the standard manner \cite{glan,egpr},
 replacing them at a
scale $\mu=500$ MeV $\cdots$ $1$ GeV by the contribution from
 resonance exchange.
Let
\be
I^r(\mu) = \sum_{R}I^R + \hat{I}^r(\mu) \; ; \; I=a_1,a_2,b \co
\label{lec3}
\ee
where the sum denotes contributions  from scalar, (axial-)vector and
tensor exchange. Our estimate
 for $I^r(M_\rho)$ consists in setting $\hat{I}^r(M_\rho) = 0$.

The quantities $I^R$ are evaluated in appendix \ref{ar}.
 The
results  of the calculation are displayed in table 2, where the
individual resonance contributions $I^R$ are listed.
Column 6 contains the sums of
those contributions  which have a definite sign.

\begin{table}[t]
\refstepcounter{table}
{Table 1: Phenomenological values {\protect\cite{glan,beg}} and source for
 the renormalized coupling
constants  $\bar{l}_i\;,i=1\ldots 6$.
The quantities $\delta_i$
in the fourth column determine the scale dependence of the $l^r_i(\mu)$
according to Eq. (\protect\ref{lec1}).
	In the text we also use $\bar{l}_\triangle = \bar{l}_6 - \bar{l}_5$, see
Eq. (\protect\ref{rr10}).}
\begin{center}
\vspace{.5cm}
\begin{tabular}{|c||r|l|r|}  \hline
$i$  & $\bar{l}_i$ &source & $16\pi^2\delta_i$  \\ \hline
  1  &$-$0.8 $\pm 1.2$  &$K_{e4},\pi\pi\rightarrow\pi\pi$ &
 1/3 \\
  2 & 5.8 $\pm 0.7$  & $K_{e4},\pi\pi\rightarrow\pi\pi$&
2/3 \\
  3    & 2.9 $\pm 2.4$  &SU(3) {\mbox{mass formulae}}     &
  $-$1/2    \\
  4    & 4.3 $\pm 0.9$  & $F_K/F_\pi$                     &
 2  \\
  5  &13.8 $\pm 1.3$ & $\pi \rightarrow e\nu\gamma$    &
 $-$1/6  \\
  6   &16.5 $\pm 1.1$&$<r^2>^\pi_V$                    &
 $-$1/3  \\
\hline
\end{tabular}
\end{center}
\end{table}
\begin{table}[t]
\refstepcounter{table}
{Table 2: Resonance contributions to the coupling constants $a_1^r,a_2^r$
and $b^r$. Column 6 contains the sums of those
 contributions which have a
definite sign.
The calculation  is presented in appendix
 {\protect\ref{ar}}}.
\begin{center}
\vspace{0.5cm}
\begin{tabular}{|r||r|r|r|r|r||r|r|}  \hline
& \multicolumn{4}{c|}{$I^R$} &&\multicolumn{2}{|c|}{$I^R$} \\ \hline
$I^r$ & $\omega$ & $\rho^0$ & $\phi$ & $A(1^{+-})$ & $\sum_R I^R$
& $S(0^{++})$&$f_2$ \\ \hline
$a_1^r$ & $-$33.2   & $-$6.1  & $-$0.1    & $0.0$ & $-$39 & $\pm 0.
8$&$\mp 4.1$ \\
$a_2^r$ &  12.5  &  2.3  & $\simeq 0$     & $-1.3$ & 13  & $\pm 1.3$&
$\pm 1.0$  \\
$b^r$   &  2.1   &  0.4  & $\simeq 0$ &  $0.7$ & 3
 & 0.0&$\pm0.5$ \\ \hline
\end{tabular}
\end{center}
\end{table}

To estimate the effects of the systematic uncertainties in the values of these
couplings, it is  useful to furthermore consider the helicity amplitudes
$H_{++}$ and $H_{+-}$ and the corresponding low-energy constants $h_\pm^r$ and
$h_s^r$,
\bea
H_{++}^\mtiny{2loops}&=& \frac{1}{(16\pi^2F^2)^2}\left\{h_+^r M^2 +h_s^r
\; s\right \} +
\cdots \co \nn
H_{+-}^\mtiny{2loops}&=& \frac{8(M^4-tu)}{s(16\pi^2F^2)^2}h_-^r +\cdots \co \nn
h_+^r &=& a_1^r +8b^r \co \;  h_s^r =a_2^r -2b^r \co \;  h_-^r=b^r \; .
\label{le19}
\eea
{}From column 6 in table 2 we obtain the central values of these couplings.
According to  experience with resonance saturation at order $E^4$, we
associate a $30\%$ uncertainty to the contributions generated by (axial-)
vector
exchange and a $100\%$ error to the contributions from scalars and from $f_2$.
Adding these errors in quadrature, we find
\be
\begin{tabular}{rrr}
$h_+^r(M_\rho)$&=&$ -14 \pm 5 \co $ \\
$h_s^r(M_\rho)$&=&  $ 7 \pm  3\co $  \\
$h_-^r(M_\rho)$&=&  $ 3 \pm  1 \; \; . $  \\
\end{tabular}
\label{le20}
\ee
Notice that tensor exchange does contribute neither to $h_+^r$ nor to
$h_s^r$, because
the coupling (\ref{are0}) is purely $D$-wave. Scalars do not affect $h_-^r$.

In Ref. \cite{bijdv}, these couplings have been determined i) from vector-meson
exchange and using nonet-symmetry, and ii) from the chiral quark model,
with the result
\bea
(h_+^r,h_s^r,h_-^r)|_{\mu = M_\rho}  =\left\{ \begin{array}{ll}
(-18,9,2) &\mbox{vector-mesons$\;$(nonet)}\\
(-12,6,2) & \mbox{chiral quark model}
\end{array}
\right.
\label{chqmc}
\eea
which agrees within the uncertainties with the values in (\ref{le20}).

This completes the determination of the parameters which occur
 at two-loop order in $\gamma \gamma \rightarrow \pi^0\pi^0$.

\setcounter{equation}{0}
\setcounter{subsection}{0}

         \section{Amplitudes and cross section to two loops\label{ac}}

         \subsection{The amplitudes: analytic  results}

         We obtain the following expression for the amplitude $A$
         to two loops,
         \be
         A = A_4 + A_6 + O(E^4) \co
         \ee
         or
         \be
         A = \frac{4 \bar{G}_\pi(s)}{s F_\pi^2} (s-M^2_\pi) + U_A + P_A +
         O(E^4).
\label{ac1}
         \ee
         The unitary part $U_A$ contains $s,t$ and $u$-channel cuts,
         and $P_A$ is a linear polynomial in $s$. Explicitly,
         \bea
         U_A &=& \frac{2 }{s F_\pi^4}\bar{G}(s) \left[ (s^2 - M_\pi^4)
         \bar{J} (s) + C(s, \bar{l}_i)  \right]
          + \frac{\bar{l}_\Delta}{24 \pi^2F_\pi^4}
         (s-M_\pi^2) \bar{J} (s)
         \nn
         &&+ \frac{(\bar{l}_2 - 5/6)}{144 \pi^2 sF_\pi^4 }(s-4M_\pi^2) \left \{
         \bar{H}(s) + 4\left[s\bar{G}(s) +2M_\pi^2
         (\stackrel{=}{G}(s) - 3
\stackrel{=}{J}(s))\right]d_{00}^2\right\}
          \nn
    &&+ \Delta_A(s,t,u) \co \nn
         \left.\right.
\label{ac2}
        \eea
         with
         \bea
         C(s,\bar{l}_i) &=& \frac{1}{48\pi^2}\left\{ 2 (\bar{l}_1 - 4/3)
      (s - 2 M_\pi^2)^2 + (\bar{l}_2
         - 5/6)(4s^2-8sM_\pi^2+16M_\pi^4)/3 \right.
         \nn
         && \left.- 3 M_\pi^4 \bar{l}_3 + 12 M_\pi^2 (s - M_\pi^2)
          \bar{l}_4 - 12s M_\pi^2 + 15M_\pi^4 \right\} \co \nn
d_{00}^2&=&\frac{1}{2}(3\cos{\theta}^
2 - 1 )  \; .
        \label{aac1}
        \eea

The loop-functions $\bar{J}$ etc.
are displayed in appendix \ref{ai}.
The quantity $\bar{G}_\pi(s)$ in Eq. (\ref{ac1}) stands for
$\bar{G}(s)$, evaluated with the physical pion mass, and
 $\theta$ denotes the scattering angle in the center-of-mass
system. The term proportional to $d_{00}^2$ in $U_A$ contributes
to $D$-waves only. For $\Delta_A$ see below.

  The polynomial
         part is
         \bea
         P_A &=& \frac{1}{(16 \pi^2 F_\pi^2)^2} [a_1 M_\pi^2 + a_2 s]
         \co
         \nn
         a_1 &=& a_1^r + \frac{1}{18} \left\{ 4 l^2 + l(8 \bar{l}_2 + 12
         \bar{l}_\Delta - \frac{4}{3}) -\frac{20}{3}\bar{l}_2+ 12
         \bar{l}_\Delta  +\frac{110}{9}\right\}\co \nn
         a_2&=& a_2^r -\frac{1}{18}\left\{l^2 +l(2\bar{l}_2 +12\bar{l}_\Delta
         +\frac{2}{3})-\frac{5}{3}\bar{l}_2 +12\bar{l}_\Delta
+\frac{697}{144} \right\} \co \nn
l&=&\hlogm \quad .
\eea

 The result for $B$ is
         \be
         B = B_6 + O(E^2) \co
         \ee
         or
         \be
         B = U_B + P_B + O(E^2) \co
         \ee
         with unitary part
         \be
         U_B = \frac{(\bar{l}_2 - 5/6)\bar{H}(s)}{288 \pi^2 F_\pi^4 s}
         + \Delta_B(s,t,u)\per
\label{ac3}
         \ee
         For the polynomial  we obtain
         \bea
         P_B &=& \frac{b}{(16 \pi^2F_\pi^2)^2} \co
         \nn
         b &=& b^r - \
\frac{1}{36} \left[ l^2 + l(2 \bar{l}_2 +
         \frac{2}{3}) -\frac{\bar{l}_2}{3} + \frac{393}{144} \right] .
\label{ac4}
         \eea
 The integrals $\Delta_{A,B} (s,t,u)$
        contain contributions from
         the two-loop box and acnode  diagrams displayed in Fig. 4.
  It turns out that  these contributions are
         very small for the cross sections below $\sqrt{s} \leq 400$
         MeV, both for $\gamma \gamma \rightarrow \pi^0 \pi^0$ ($0.1\%$ at
 400 MeV) and for the
crossed channel $\gamma \pi^0 \rightarrow \gamma \pi^0$ ($1.5\%$ at 400
 MeV).
Therefore,
         one obtains a rather compact and accurate representation of
         the two-loop amplitudes by simply setting $\Delta_{A,B} =
          0$ in $U_{A,B}$\footnote{An analogous result holds for
         the elastic $\pi \pi$ scattering amplitude, which contains
         unitarity contributions with $t$- and $u$-channel cuts which
         are negligible below $E \simeq 500$ MeV for $S$-waves.}.

         \subsection{The cross section $\gamma \gamma \rightarrow \pi^0
         \pi^0$}

         In Fig. 5 we display the data for the cross section
$\sigma(s ; |\cos \theta| \leq Z = 0.8)$ as determined in the
 Crystal Ball experiment \cite{cball}. They are shown as a function of
the center-of-mass energy  $E = \sqrt{s}$.
 The solid line denotes the two-loop result,
         evaluated\footnote{We use $F_\pi = 93.2$ MeV,
         $M_\pi = M_{\pi^0} = 135$ MeV, unless stated otherwise.}
         with the amplitudes (\ref{ac1})-(\ref{ac4}). For the
 low-energy constants
$a_1^r,
a_2^r$ and $b^r$ we have used the values from column 6 in
table 2,
 and the values of
         $\bar{l}_i$ are the ones displayed in table 1.
         Shown is furthermore, with a dashed line, the one-loop result
\cite{bico},
 obtained  by setting $U_{A,B}=P_{A,B}=0$,
 see also appendix \ref{ao}.
        Finally, the dash-dotted
         lines display the result of a dispersive analysis (Fig. 23 in Ref.
\cite{pehan}).
         In that calculation, use was made of the $I=0,2$ $S$-wave
$\pi \pi$ phase shifts from Ref.
         \cite{schenk}
         (these phase shifts satisfy  constraints imposed
         by Roy-type equations  \cite{sternpp}).

The two-loop result thus  agrees well with the data and with the dispersive
analysis of Pennington \cite{pehan} in the low-energy region.

         We found it interesting to see which contributions are
         responsible for the increase in amplitude and cross section
         near the threshold. In Fig. 6, we display with a
         solid line the cross section, evaluated at $\Delta_{A,B} =
           0$ and without resonance exchange. [The change
         compared to the full result (solid line in Fig. 5)
         is 0.2 nb at $E =$ 400 MeV and thus negligible.] The dashed
         line corresponds to $\bar{l}_i = 0$, and the dash-dotted
         line is obtained by setting $\bar{l}_1 = \bar{l}_3 = 0$. We
         conclude that the increase is due to $\bar{l}_2$, $\bar{l}_4$
         and $\bar{l}_\Delta=\bar{l}_6 - \bar{l}_5$. To make this statement
more quantitative, we note that the dependence
   of the cross section on the $\bar{l}_i$-values
        can been summarized with the expression
 \bea
         \sigma^\mtiny{2loops}(s) & \simeq & N\sigma^\mtiny{1loop}(s) \co \nn
N&=&1 + (-5.8+5.0\bar{l}_1 + 4.9\bar{l}_2
                 - 0.2\bar{l}_3
                 + 5.4\bar{l}_4 + 3.7\bar{l}_\Delta)\cdot10^{-2} \nn
                   &=&1-0.058-0.040+0.283-0.005+0.232+0.100 \nn
                   &\simeq& 1.51
\label{ac5}
\eea
        which is accurate to a few percent up
         to 450 MeV.
The analogous expression
        for the helicity amplitude $H_{++}$ at the physical
        threshold $s=4M_\pi^2$ reads
        \bea
   H^\mtiny{2loops}_{++} & = & N H_{++}^\mtiny{1loop} \co \nn
 N&=&1 + (2.5 + 0.6\bar{l}_1
 +1.2\bar{l}_2 - 0.2\bar{l}_3 + 2.7\bar{l}_4 + 2.7\bar{l}_{\Delta})\cdot
10^{-2}\nn
&=& 1+0.025-0.005+0.068-0.006+0.114+0.073\nn
&\simeq&1.27 \co
\label{ac6}
\eea
with
\be
M_\pi^2 H_{++}^\mtiny{1loop}  = 5.8\cdot 10^{-2}.
\ee
The contributions from $\bar{l}_1,\bar{l}_2$  are
 $\pi \pi$ rescattering effects. They amount to a $24\%$ increase in
the cross section (out of $51\%$) and to $6\%$ in $H_{++}$ (out of $27\%$).
 The renormalization of $F_\pi$ (contribution from
$\bar{l}_4$) amounts to a $23\%$ increase in $\sigma$.

         \subsection{The amplitudes: numerical results}

         To get more insight into the characteristics of the two-loop
         corrections, we display in Fig. 7 the real and
         imaginary part of the helicity amplitudes
 $\pm 10^2M_\pi^2H_{+\pm}$ at $t=u$. The
solid line shows $10^2M_\pi^2H_{++}$. It incorporates all contributions
except $\Delta_{A,B}$. The dashed line is the same amplitude for the
 one-loop
case, and the dash-dotted line is for  $-10^2M_\pi^2H_{+-}$ with
 the same input as  for
the solid line. The
         curves start at $E=2 M_\pi$, and the crosses refer to the
 center-of-mass energy of the $\pi^0 \pi^0$ system in 100 MeV steps.
 The amplitude $H_{++}$ changes very
         rapidly just above threshold and is nearly purely
         imaginary in the region 350 MeV $\leq E \leq 400$ MeV.
 As expected, the amplitude $H_{+-}$ which
         starts out with a $D$-wave term is very small at low energies.
         Resonance exchange adds to $H_{++}$ a positive real
         part, thus increasing the cross section below $\sim 400$ MeV
         and decreasing it above this energy.

         In Fig. 8, we display the quantity $10^2 M^2_\pi
         H_{++}$ at $t=u$ as
         a function of $s/M^2_\pi$.
 Above the threshold $s = 4
         M^2_\pi$, the modulus is shown. The solid (dashed) line
         denotes the full two-loop (one-loop) result. While the
         two-loop contribution to the modulus is below 30\% in the
         threshold region, it modifies $H_{++}$ substantially (percentage-wise)
at
         $s=0$, where the amplitude is small, see also the discussion below.
Furthermore, we note that the zero at $s =
         M^2_\pi$, which occurs in the one-loop approximation Eq.
         (\ref{ao1}), is only slightly modified by the loop corrections.
         Finally, we display with $\diamond$ (+) the modulus of the
         $S$-wave projected part of $H_{++}$, taken from Fig. 19 (23) in
 Ref. \cite{pehan}.

\subsection{Error estimates and range of validity of the chiral
representation}
The uncertainty in the amplitude has two sources.
Firstly, the low-energy constants $\bar{l}_i,h_\pm^r$ and $h_s^r$ used above
contain certain errors. For the $\bar{l}_i$, these are displayed in
 table 1. The systematic errors in the low-energy couplings at order $E^6$ have
been estimated in the previous section, see Eq. (\ref{le20}).
 Secondly, we are
concerned here with an expansion in powers of the quark masses and of the
external momenta. Higher order terms in this expansion  (three  loops and
beyond)
will change the cross section accordingly.

We discuss first the effect of the uncertainty in the low-energy
constants and concentrate for simplicity on $h_{\pm}^r$ and $h_s^r$.
In Fig. 9 we show the variation of the cross section according to the error
estimates in
Eq. (\ref{le20}). The calculation is done at $\triangle_{A,B}=0$.
The dashed lines  embrace the region generated by
assigning all possible combinations of signs to the systematic errors
in the couplings $h_\pm^r$ and $h_s^r$ according to Eq. (\ref{le20}).
 The dash-dotted line corresponds to the central value in
(\ref{le20}).

It is clearly seen that, below 400 MeV, the uncertainties in $h_\pm^r$ and in
$h_s^r$ do not matter. (Since we estimate the couplings with resonance
saturation, this is a reformulation of earlier findings
\cite{ko1}-\cite{ba2loop}.)
Varying the scale at which resonance saturation is assumed
between 500 MeV and 1 GeV results also in a negligible change
  in the cross section  below $E=$ 400 MeV.
 Beyond this energy, the  uncertainty
 becomes  more pronounced.
 Because the contribution from $H_{+-}$
is tiny (see Fig. 7), only $h_+^r$ and $h_s^r$ really count.
 One might
thus be tempted to extract these couplings from more accurate data
 in the range $E=(400-600)$ MeV. This
would be interesting, because $h_+^r$ determines the difference of the
 electric and
 magnetic polarizabilities of the neutral pion at two-loop order,
 see below.
However, in this energy range where $s\simeq (9-20)M_\pi^2$, $h_s^r$ is much
more important than $h_+^r$. It will, therefore, be rather
difficult to extract $h_+^r$ reliably in this manner \cite{dohod}.
 On the other hand, it would be
interesting to perform a combined analysis of the two related processes
$\gamma \gamma \rightarrow \pi^0 \pi^0$ and $\eta \rightarrow \pi^0 \gamma
\gamma$ \cite{wzbijn} in the framework of $SU(3)_L \times SU(3)_R$
 in order to obtain maximal
information on the low-energy coupling constants which enter these
amplitudes\footnote{We thank J. Bijnens, M. Knecht and J. Stern for
 pointing this out to us.}.

Turning now to the corrections from yet higher orders, we use the fact that
$\sigma^\mtiny{2loops}/\sigma^\mtiny{1loop} \simeq (1+\epsilon)^2$ with
$\epsilon \simeq 0.25$ and estimate  $\sigma/\sigma^\mtiny{1loop}
 \simeq (1-\epsilon)^{-2}$. This amounts to a $15-20\%$ uncertainty in
the two-loop result below $400$ MeV. At higher energies, the error in the cross
section is more difficult to assess. It may well turn out, however, that a
more precise determination of the low-energy couplings leads to the
 conclusion that the chiral
representation of the amplitude at the two-loop level is even valid
 up to $E=(600-700)$ MeV in this
channel.

\setcounter{equation}{0}
\setcounter{section}{7}
\setcounter{subsection}{0}

\section{Compton scattering and pion polarizabilities}

The amplitudes $A$ and $B$ are analytic functions of $s$ and $t$. At $s\leq 0$,
they describe Compton scattering,
\be
\gamma(q_1) \pi^0(p_1) \rightarrow \gamma(q_2)\pi^0(p_2) \per
\label{cs1}
\ee
We discuss this reaction in the present section, where we also work out the
(neutral) pion polarizabilities at next-to-leading order.

\subsection{Compton scattering}
The cross section $\gamma \gamma \rightarrow \pi^0 \pi^0$ receives a
substantial
correction near threshold due to $\pi\pi$ final-state interactions -- which are
absent in Compton scattering. Are then the two-loop contributions small
in this
channel?  Fig. 8 shows that this is not the case: in the
one-loop approximation, the amplitude $H_{++}$ is one order of magnitude larger
in the $\gamma \gamma \rightarrow \pi^0 \pi^0$ channel than at
Compton threshold.
Therefore, even tiny corrections
in $\gamma \gamma \rightarrow \pi^0 \pi^0$ may appear large in Compton
scattering \cite{ba2loop}. In
 Fig. 10 we display the  cross section
 $\sigma^{\gamma \pi^0\rightarrow \gamma \pi^0}$
  as a function of the
center-of-mass energy $E_{\gamma\pi}$. The solid line shows the result of the
two-loop calculation and the dashed line displays the one-loop
approximation.
They differ by one order of magnitude already near threshold.
This is mainly due  to the effect of the low-energy constant $h_-^r$ in
$H_{+-}$
(omega-exchange in the language of resonance saturation \cite{ba2loop}).
 Putting $H_{+-}$ to zero results in  $\sigma^{\gamma \pi^0 \rightarrow
\gamma \pi^0}=0.7 \; \mbox{nb}$ at $E_{\gamma\pi}=350$ MeV
 (dotted line).
Purely two-loop effects however also change the cross section by
roughly
a factor of two at $E_{\gamma\pi}=350$ MeV (dash-dotted line,
 evaluated  at $h_{\pm}^r=h_s^r=0$).

In summary, the Compton amplitude is tiny at leading order, and it is therefore
 rather unstable against the corrections generated by
higher order terms.

         \subsection{Pion polarizabilities}

For a composite system it is customary to include the
electric and magnetic polarizabilities among the fundamental
parameters--such as the electric charge, the magnetic moment
and the mass--characterizing the low-energy limit of the coupling
with the photon in the Compton amplitude. Hadrons are no
exception, hence the theoretical description of their dynamics can
be tested through the experimental determination of the hadron
polarizabilities \cite{revpol}. To set notation, we first  consider
          Compton scattering for charged pions,
         \be
         \gamma(q_1)  \pi^+(p_1) \rightarrow \gamma(q_2)
         \pi^+(p_2)\co
         \ee
         in the laboratory system  $p_1^0=\hmpp$. (In order to simplify the
notation, we use the symbol $M_\pi$ to denote both the charged and the neutral
pion mass.) The electric
         $(\bar{\alpha}_{\pi})$ and magnetic $(\bar{\beta}_{\pi})$
         polarizabilities are obtained by expanding the Compton amplitude at
threshold,
          \bea
         T^C &=& 2 \left[ \vec{\epsilon}_1 \cdot \vec{\epsilon}_2 \,\! ^\star
 \left(
         \frac{\alpha}{\hmpp} - \bar{\alpha}_{\pi} \omega_1 \omega_2
         \right)- \bar{\beta}_{\pi} \left(\vec{q}_1 \times \vec{\epsilon}_1
         \right) \cdot \left(
         \vec{q}_2 \times \vec{\epsilon}_2 \,\! ^\star \right)
          + \cdots \right]
         \eea
         with $q_i^\mu = (\omega_i, \vec{q}_i)$. In terms of the
         helicity components  (\ref{ao0}), one has
         \be
         \bar{\alpha}_{\pi} \pm \bar{\beta}_{\pi}=-\frac{\alpha}{\hmpp}
\bar{H}^C_{+\mp}(s=0,t=\mppq) \co
         \ee
         where the bar denotes the amplitude with the Born term
removed\footnote{
In $\bar{H}^C_{+-}$ first set $t=\mppq$, then $s\rightarrow 0$. We use the
Condon-Shortley phase convention.}. For { neutral} pions, one uses the
         analogous definition,
         \be
         \bar{\alpha}_{\pi^0} \pm \bar{\beta}_{\pi^0}=\frac{\alpha}{M_\pi}
             H_{+\mp}(0,M_\pi^2)\co
         \ee
or, in terms of $A$ and $B$,
          \bea
\bar{\alpha}_{\pi^0} &=&\frac{\alpha}{2M_\pi}(A+16M_\pi^2B)|_{s=0,t=M_\pi^2}\co
\nn
  \bar{\beta}_{\pi^0}&=&-\frac{\alpha}{2M_\pi} A|_{s=0,t=M_\pi^2}.
\eea
Below we also use the notation
\bea
(\alpha \pm \beta)^C &=&   \bar{\alpha}_{\pi} \pm \bar{\beta}_{\pi}
\co \nn
(\alpha \pm \beta)^N &=&   \bar{\alpha}_{\pi^0} \pm \bar{\beta}_{\pi^0} \; \; .
\eea
An unsubtracted forward dispersion relation for the amplitude $B$ gives
 with (\ref{na10})
\be
\apbn = \frac{M_\pi}{\pi^2}\int_{4M_\pi^2}^\infty
\frac{ds'}{(s'-M_\pi^2)^2}\sigma_\tot^{\gamma \pi^0}(s')\co
\ee
and analogously for the charged channel.

\subsection{Data on pion polarizabilities}

There exist up to now two determinations of charged pion polarizabilities via
measurement of the Compton amplitude. At Serpukhov \cite{serpukov1}, radiative
pion-nucleus scattering $\pi^-A\rightarrow \pi^-\gamma A $ has been used. Here
the incident pion scatters from a virtual photon in the Coulomb field of the
nucleus. In the  pion production process $\gamma p
\rightarrow \gamma \pi^+n$  examined at the Lebedev Institute \cite{lebedev},
the incoming photon scatters from a virtual
pion. Analyzing the data with the constraint  $\apbc=0$
gives\footnote{We express the polarizabilities in units of $10^{-4}\mbox{fm}^3$
throughout.}
\bea
\ambc  =\left\{ \begin{array}{ll}
13.6 \pm 2.8 & \cite{serpukov1} \\
\; \; \; 40 \pm 24 & \cite{lebedev} \per
\end{array}
\right.
\label{cs3}
\eea
The Serpukhov data have been analyzed  also relaxing the constraint
 $\apbc = 0$,
with the result
\bea
 \apbc&=& \; \; 1.4 \pm 3.1 (\mbox{stat.}) \pm
2.5 (\mbox{sys.}) \; \;  \cite{serpukov2} \co \nn
 \ambc&=& 15.6 \pm 6.4 (\mbox{stat.}) \pm 4.4 (\mbox{sys.})
\; \; \cite{serpukov2} \per
\label{cs4}
\eea
Here we have converted the  value quoted for $\bar{\beta}_{\pi}$ into
 a number for
$(\alpha - \beta)^C$, adding the errors in quadrature.

Furthermore, also the  process $\gamma \gamma \rightarrow \pi \pi$ may be used
to
obtain information on the polarizabilities. Since in this case the
 amplitude at low
energies is mainly sensitive to $S$-wave scattering, only  $\ambcn$  can be
determined from the presently available
\cite{dcharged,cball} data. In
Ref. \cite{kaloshin}, unitarized $S$-wave amplitudes have beeen constructed,
 which contain   $\ambcn$ as adjustable parameters. A simultaneous fit to
Mark II  and Crystal Ball data gives
\bea
  \ambc &=& \; \;\; 4.8 \pm 1.0\; \; \cite{kaloshin} \co \nn
 \ambn&=& -1.1 \pm 1.7 \; \;  \cite{kaloshin} \co
\label{cs5}
\eea
where we have taken into account that the definition of the
polarizabilities in \cite{kaloshin} is $4\pi$ larger than the one used here,
see \cite{kaloshinold}, Eq. 1.

The value (\ref{cs5}) for $\ambc$ is consistent with Refs.
 \cite{lebedev,serpukov2}  within $1\frac{1}{2} $ standard deviations, but not
consistent with \cite{serpukov1}.
The large relative error in $\ambn$ reflects the fact that the
threshold amplitude $\gamma \gamma \rightarrow \pi^0 \pi^0$ is quite
insensitive to large relative changes at the Compton threshold,
 as we discussed above
 (see also Fig. 10 in \cite{dohod}).
The determination of
 $\ambcn$ from $\gamma \gamma \rightarrow \pi \pi$ furthermore suffers from
 uncertainties which we find difficult to estimate in the approach
used by Kaloshin and Serebryakov \cite{kaloshin}, which does not provide a
 systematic way to control the inherent uncertainties in the model amplitude
used to fit the data. It might be interesting to merge dispersion
 relations and CHPT
  at next-to-leading order in the chiral
expansion. This method \cite{higgs,dohod}, which does provide a control on the
approximations
made, would then allow for an experimental
determination of
 $\ambcn$
 at order $E$, requiring, however, a two-loop evaluation of
 $\gamma \gamma \rightarrow \pi^+ \pi^-$.

In Ref. \cite{bound}, the bound $|\bar{\alpha}_{\pi^0}|< 35$ has been obtained
from a study of the $e^+e^-\rightarrow \pi^0\pi^0\gamma$ reaction.

Finally,  information on the charged pion polarizabilities
may be obtained from  $\gamma \gamma \rightarrow \pi^+\pi^-$ data in the
following manner \cite{bbgm0}. Both the one-loop expression for the transition
amplitude and the leading-order expression for $\bar{\alpha}_\pi$ and
$\bar{\beta}_\pi$ contain the low-energy constant $\bar{l}_\triangle$ as the
only free parameter. Extracting it from a fit to the cross section
 then determines $\bar{\alpha}_\pi$ and
$\bar{\beta}_\pi$ at this order. The result \cite{bbgm0}
 $\bar{l}_\triangle = 2.3 \pm 1.7$  agrees within the error with the
value  $\bar{l}_\triangle = 2.7$ used in the present work--the corresponding
numerical values for the leading-order expressions of $(\alpha \pm \beta)^C$
therefore also agree.  The cross section in the threshold
region is dominated by the Born term contribution and is, therefore,
rather insensitive to $\bar{l}_\triangle$ \cite{bico}. This is the main
reason for the large uncertainty in this determination of $\bar{l}_\triangle$.

\subsection{Chiral expansion of  $\bar{\alpha}_{
\pi}$ and $\bar{\beta}_{\pi}$ }

The one-loop result is
\bea
\apbc &=& 0 \co \nn
\apbn &=& 0 \co \nn
\ambc &=& \frac{\alpha \bar{l}_\triangle}
{24 \pi^2M_{\pi}F^2} = 5.3 \co \nn
\ambn &=&-\frac{\alpha}{48\pi^2M_{\pi}F^2} = -1.0 \; \; .
\label{cs6}
\eea
Here we have identified $F$ with the physical value of the
pion decay constant, and we have used the charged pion mass to evaluate
of $\ambc$.
It is straightforward to determine from the amplitudes given in the previous
section the neutral pion polarizabilities
to two loops. The numerical results are
displayed in table 3.
\begin{table}[t]
\protect
\begin{center}
\caption{Neutral pion polarizabilities to two loops
 in units of $10^{-4}\mbox{fm}^3$. The contribution due to
chiral logarithms, listed in the fifth column with bracketed numbers, is
 included in the two-loop
result quoted in column four.}

\vspace{1em}
\begin{tabular}{|r||r|r|r|r||r|r|} \hline
 &$O(E^{-1})$&\multicolumn{3}{c||}{$O(E)$}  & &
 \\ \cline{2-5}
  & 1 loop & $h_\pm^r$&2 loops&chiral logs&total&uncertainty
\\ \hline
$\apbn$&0.00& 1.00 & 0.17 &[0.21]&$\simeq 1.15$&
$ \pm 0.30$\\ \hline
$\ambn$&$-$1.01 & $-$0.58 &$-$0.31
 &[$-$0.18]&$\simeq -1.90$& $\pm 0.20$
\\ \hline
$\bar{\alpha}_{\pi^0}$&$-$0.50 & 0.21 & $-$0.07 &[0.01]& $\simeq-$0.35&
 $\pm 0.10$ \\ \hline
$\bar{\beta}_{\pi^0}$&0.50 & 0.79 &0.24 &[0.20]& $\simeq$1.50&
$ \pm 0.20$ \\ \hline
\end{tabular}
\end{center}
\end{table}
 The second column contains the contribution at order
$E^{-1}$, and the third to fifth  columns display the terms of order
$E$. The total values are given in column 6.
 (The two-loop
contribution $\apbn = 0.18$ reported earlier \cite{gaap93} and quoted in
Ref. \cite{ba2loop} corresponds to slightly different values of $\bar{l}_1$ and
$\bar{l}_2$.)
Finally, our  estimate
of the errors is shown in the last column. These are obtained in the same
manner as the ones for the couplings $h_\pm^r$ and $h^r_s$ in Eq. (\ref{le20}).
 We have not
considered correlations in these uncertainties, which do also not  contain
effects from higher orders in the quark mass expansion.

The contribution
from the chiral logarithms present in the low-energy expansion of the
polarizabilities deserves a comment. As we discussed earlier, the
 $\ln^2M_\pi^2/\mu^2$ terms occur in
a particular combination which is dictated by the general structure of the
renormalized amplitude,
\bea
(\alpha \pm \beta)^N_\mtiny{2loops} &=& C_\pm L_\chi+\cdots \co \nn
 L_\chi &=& \frac{\alpha M_{\pi}}{(16\pi^2F_\pi^2)^2}\ln M_\pi^2/\mu^2\left\{
\ln M_\pi^2/\mu^2 +2 \bar{l}_2\right\} \; \; .
\eea
Here the ellipsis denotes further single logarithms and terms of order $M_\pi$,
and $C_\pm$ are Clebsch-Gordan coefficients.
 These terms are
potentially very large,
\be
L_\chi  = -1.14 \cdot 10^{-4} {\mbox{fm}}^3
\ee
at $\mu = M_\rho$.
 It turns out  that $C_+$ is small, whereas $C_-$ even vanishes.
  We have listed the sum
of the $\ln^2M_\pi^2/\mu^2$ and $\ln M_\pi^2/\mu^2$
terms at the scale $\mu=M_\rho$ in the fifth column of
 table 3--these contributions are
 included in the two-loop
result quoted in column four.

The low-energy constants determined in \cite{bijdv} give for the contributions
from $h_\pm^r$
\bea
(\bar{\alpha}_{\pi^0},\bar{\beta}_{\pi^0})  =\left\{ \begin{array}{ll}
(0.0,0.72) &\mbox{vector-mesons$\;$ (nonet)}\\
(0.0,0.50) & \mbox{chiral quark model}
\end{array}
\right.
\label{chqmp}
\eea
The corresponding entries in column 3 of table 3 are slightly different than
the ones from vector-exchange in Eq. (\ref{chqmp}), because we do not use the
nonet-assumption here and include in addition axial-vector
exchange. The slight discrepancy with the chiral quark model
prediction is not serious, because the systematic uncertainties in that
framework
are very difficult to assess.

Turning now to a comparison with the data, we note  that the two-loop result
for $\ambn$ agrees within the error bars with the value found by Kaloshin and
Serebryakov \cite{kaloshin}. As for the charged pion case, the
 complete  expression at order $E$ is
not yet available.
 The chiral logarithms which  occur at this  order
in the low-energy expansion can in principle contribute
 substantially also here.
Therefore,  to compare the chiral prediction with the data, a full
two-loop calculation is required \cite{buergi}.

\setcounter{equation}{0}
\setcounter{subsection}{0}

\section{Comparison with dispersion relations\label{cd}}

In this section we  compare in some detail the
chiral expansion with the dispersive calculation carried out by Donoghue and
Holstein
 \cite{dohod}.

\noindent
Consider the $S$-wave amplitude
\be
F(s)=\frac{1}{4\pi}\int d\Omega H_{++}(s,t) \; .
\ee
We find from the two-loop representation given above
\bea
F^\CHPT&=&\frac{2}{sF_\pi^4}\bar{G}_\pi(s)\left\{2F_\pi^2 (s-M_\pi^2) +
(s^2-M_\pi^4)\bar{J}(s)
+C(s,\bar{l}_i)\right\}
\nn
 &&+\frac{\bar{l}_\Delta}{24\pi^2F_\pi^4}(s-M_\pi^2)\{\bar{J}(s)
-\frac{1+l}{16\pi^2}\}
+P_F +\Delta_F\co \nn
P_F&=& \frac{1}{(16\pi^2F_\pi^2)^2}[f_1M_\pi^2 +f_2s]\co \nn
f_1&=&h_+^r-\frac{1}{9}\left\{2l+\frac{8}{3}\bar{l}_2
-\frac{47}{72}\right\}\co\nn
f_2&=&h_s^r +\frac{1}{18}\left\{\frac{4}{3}\bar{l}_2
-\frac{19}{9}\right\}\co \nn
l&=&\hlogm \co
\label{cw1}
\eea
where $C(s,\bar{l}_i)$ is displayed in Eq. (\ref{aac1}), and where $\Delta_F$
 is the $S$-wave contribution from $\Delta_{A,B}$.
In the region $2M_\pi \leq E \leq 400$ MeV, the polynomial $P_F$ contributes
very little to the amplitude. For the comparison with the dispersive approach
in this region
we  therefore drop it, together with $\Delta_F$.

In the
simplest version of their analysis, Donoghue and Holstein write
\bea
F^\DISP&=&\frac{4\bar{G}_\pi(s)}{3sF_\pi^2}\left\{
(2s-M_\pi^2)D_
0^{-1}(s) +(s-2M_\pi^2)D_2^{-1}(s)\right\}\nn
&+&\frac{\bar{l}_\Delta}{36\pi^2F_\pi^2}\left\{D_0^{-1}(s) -D_2^{-1}(s)\right\}
\co
\label{cw2}
\eea
where $D_I^{-1}$ is the Omn\`es function
\be
D_I^{-1}=\frac{1}{1-k_Is -t_I^\CA16\pi\bar{J}(s)}\co
\ee
with
\bea
k_0&=&\frac{1}{25M_\pi^2}\scs k_2=-\frac{1}{30M_\pi^2} \co \nn
t_0^\CA&=&\frac{2s-M_\pi^2}{32\pi F_\pi^2}\scs
t_2^\CA=-\frac{s-2M_\pi^2}{32\pi F_\pi^2}\; .
\eea
Expanding $D_I^{-1}$  and keeping terms of the same order as in
$F^\CHPT$, we find (we count $k_I$ as order $E^0$)
\bea
F^\DISP&=&
\frac{2}{sF_\pi^4}\bar{G}_\pi(s)\left\{2F_\pi^2 (s-M_\pi^2) +
(s^2-M_\pi^4)\bar{J}(s) +  C^\DISP
\right\} \co \nn
&&+\frac{\bar{l}_\Delta}{24\pi^2F_\pi^4}\left\{(s-M_\pi^2)\bar{J}(s)
+\frac{2}{3}F_\pi^2s(k_0-k_2)\right\} +O(E^4) \co \nn
 C^\DISP&=& \frac{2F_\pi^2s}{3}[k_0(2s-M_\pi^2) +k_2(s-2M_\pi^2)]
\;.
\label{cw3}
\eea
The two representations (\ref{cw1}) and (\ref{cw3}) give very similar cross
sections up to $E\simeq 400$ MeV.
This is at first surprising, because the
polynomial $C(s,\bar{l}_i)$ in the chiral representation
(\ref{cw1})
contains rescattering effects which are algebraically quite different from
 $C^\DISP$
(e.g., the leading terms $\simeq s^2$ differ by more than a factor of 3).
 The  polynomial multiplying $\bar{l}_\Delta$ is also
different in the two representations. The combined effect of these two
differencies is  that the $S$-wave amplitude
(\ref{cw3}) agrees numerically quite well with (\ref{cw1}).
[Notice that Fig. 3 in Ref. \cite{dohod} which displays
the cross section according to Eq. (\ref{cw2}) is not correct
 \cite{kambor}.]

Donoghue and Holstein then refine their representation (\ref{cw2}) by adding
contributions from resonance exchange. Their final result for the cross section
agrees very well with our representation below $E=400$ MeV, see
Fig. 11.
 There we display with a solid line the two-loop result. The
dashed line is the result of Donoghue and Holstein (Fig. 2 in \cite{dohod}).
 The two
representations differ in the threshold region, because $M_\pi$ is identified
with the charged pion mass by these authors.

There {\it{are}} differences in the two representations, though. First, in the
dispersive method, higher order terms are partially summed up.
 We consider the fact that
the cross sections agree as an indication that yet higher orders in the
chiral expansion do not affect the amplitude in the threshold region very much.
Secondly,  CHPT reveals that the amplitude contains chiral
logarithms, generated by pion loops. All of these effects are not
incorporated in the dispersive analysis of Ref. \cite{dohod}. To illustrate,
 consider the amplitude
$F$ at the Compton threshold, where it determines the difference
of the electric and  magnetic polarizabilities,
\be
 \ambn=
\frac{\alpha}{M_\pi}F(0)\; .
\ee
Numerically, the chiral logarithms amount to a
$18\%$ correction to the leading-order term
$\ambn = -1.01$,
 see table 3. The result
$\ambn =-1.76$ quoted in Ref. \cite{dohod} corresponds to the
one-loop contribution and to vector exchange alone and therefore differs from
our value
$\ambn \simeq -1.90$. In $\bar{\beta}_{\pi^0}$,
axial-vectors do not contribute. The logarithms amount to  $0.20$ in the final
result
$\bar{\beta}_{\pi^0} \simeq 1.50$  which differs by $20\%$ from the value
$\bar{\beta}_{\pi^0}=1.26$ in \cite{dohod}.

\newpage

\section{Summary and conclusion\label{sc}}

\begin{enumerate}
\item At leading order in the chiral expansion, the amplitude for $\gamma
\gamma \rightarrow \pi^0 \pi^0$ is generated by one-loop graphs
 \cite{bico,dhlin}. In the case of
$SU(2)_L\times SU(2)_R\times U(1)$ considered here, it
involves the pion decay constant and the pion mass as the only parameters.
The corresponding cross section deviates from the data and from dispersive
calculations already near  threshold.

\item The neglected terms in this  calculation are related  to
 $\pi \pi$ final-state interactions and to
 three new
low-energy constants $h_\pm^r$ and $h_s^r$ which occur at
 order $E^6$ in the effective action.

\item To investigate these corrections, we have evaluated the
 next-to-leading order terms in the chiral expansion (two-loop diagrams)
and have
 estimated  the new couplings   in the standard manner   \cite{glan,egpr}
 by resonance saturation ($J^{PC}=0^{++},1^{- -},1^{+-},2^{++}$).

\item The improved cross section agrees rather well with the data and with
dispersion
theoretic calculations at and also substantially above the threshold region,
see Fig. 5 and Fig. 11. The enhancement in the cross section is mainly
 due to $\pi \pi$
rescattering and to the renormalization of the pion decay constant.

\item
The two-loop corrections
are not unduly large--their size is similar to the corresponding
next-to-leading
order correction in the isospin zero $\pi \pi$ scattering amplitude
 \cite{glan} and in the scalar form factor of the pion \cite{gmff}.

\item The couplings $h_\pm^r$ and $h_s^r$ contribute with a
negligible amount below $E=400$ MeV \cite{ko1}-\cite{ba2loop}.
 Above this energy, the inherent
uncertainty in $h_s^r$ becomes more important (Fig. 9). The influence
of $h_\pm^r$ is quite small also in the region
 $400$ MeV$ \leq
E \leq 600$ MeV.

\item
The amplitude for the crossed reaction $\gamma \pi^0 \rightarrow \gamma
\pi^0$ is small at the threshold $E_{\gamma \pi} = M_\pi$. As a result of
this, the one-loop representation is substantially distorted by the
next-to-leading order terms, although there are no final-state interactions in
this case. The dominant effect is due to $h_-^r$ (omega exchange $\gamma \pi^0
\rightarrow \omega \rightarrow \gamma \pi^0$ in the language of resonance
saturation \cite{ba2loop}).

\item
The quark mass expansion of the pion polarizabilities
 $\bar{\alpha}_{\pi^0}$ and $\bar{\beta}_{\pi^0}$ contains
  chiral logarithms
$\sim M_\pi \ln^2 M_\pi$ and $\sim M_\pi \ln M_\pi$
 which contribute substantially to
 $\bar{\alpha}_{\pi^0} \pm \bar{\beta}_{\pi^0}$, although their effect is
suppressed by small
Clebsch-Gordan coefficients. The effect of the low-energy constants $h_\pm^r$
on the value of  $\bar{\alpha}_{\pi^0} \mp \bar{\beta}_{\pi^0}$ is large.
It will presumably be difficult to extract these couplings from low-energy
$\gamma \gamma \rightarrow \pi^0 \pi^0$ data alone and to determine in
this manner
the polarizabilities at two-loop order \cite{dohod}.

\item
The DAFNE facility \cite{handbooko,handbookn} will have the opportunity to
test the chiral predictions
at next-to-leading order in much more detail than  is possible with  present
data.
\end{enumerate}

\vspace{3cm}

\noindent
{\bf Acknowledgements}\\
It is a pleasure to thank for useful discussions with or support from
Ll. Ametller, D. Babusci,
J. Bijnens, A. Bramon, G. D'Ambrosio, J.F. Donoghue, G. Ecker, M. Egger,
H. Genz, O. H\"anni, A. Held, F. Jegerlehner,
M. Knecht, I. Ku\v{s}ar, H. Leutwyler,
L. Maiani, G.J. van Oldenborgh, G. Pancheri, M. Pennington, J. Stern
and G. Weiglein.
One of the authors (MES) thanks the Magnus Ehrnrooth Foundation for a grant.

\newpage

\newcounter{zahler}
\renewcommand{\thesection}{\Alph{zahler}}
\renewcommand{\theequation}{\Alph{zahler}.\arabic{equation}}

\setcounter{zahler}{0}

\appendix

\setcounter{equation}{0}
\addtocounter{zahler}{1}
\section{Decomposition of $V^{\mu \nu}$ \label{ad}}

         Here we briefly discuss the correlator
         \be
         V_{\mu \nu} = i \int dx e^{-i(q_1x + q_2y)} <\pi^0 (p_1)
         \pi^0(p_2) \mbox{out} \mid Tj_\mu(x) j_\nu (y) \mid 0 >.
         \ee
         We decompose $V^{\mu \nu}$ into Lorentz and parity
         invariant amplitudes
         \bea
         V_{\mu \nu} &=& V_0 g_{\mu \nu} + V_1 q_{1\mu} q_{1\nu} + V_2
         q_{1\mu}q_{2\nu} + V_3 q_{1\mu} \Delta_\nu
         \nn
         &&+ V_4 q_{1\nu} q_{2\mu} + V_5 q_{1 \nu} \Delta_\mu + V_6
         q_{2\mu} q_{2\nu} + V_7 q_{2\mu} \Delta_\nu
         \nn
         &&+ V_8 q_{2\nu} \Delta_{\mu} + V_9 \Delta_\mu \Delta_\nu \co
         \nn
         \Delta_\mu &=& (p_1 - p_2)_\mu
         \nn
         V_i &=& V_i (s, \nu) \sem i = 0, \ldots, 9\co
          \nn
         s&=&(q_1+q_2)^2\scs t=(p_1-q_1)^2\scs u=(p_2-q_1)^2 \co
          \nn
         \nu&=&t-u \; .
         \label{ad1}
         \eea
         From Bose symmetry
         \be
         V_{\mu \nu} (\Delta, q_1, q_2) = V_{\mu \nu} (- \Delta, q_1,
         q_2) = V_{\nu \mu} (\Delta,q_2, q_1)
         \ee
         we find
         \bea
         V_i(s, \nu) &=& V_i (s,-\nu)\sem i = 0,1,2,4,6,9
         \nn
         V_i(s,\nu) &=& -V_i (s,-\nu)\sem i = 3,5,7,8
         \eea
         and
         \be
         V_6 = V_1\;,\; V_5 = -V_7\;,\; V_8 = - V_3 \per
         \ee
         The Ward identity
         \be q^\mu_1 V_{\mu \nu} = 0
         \ee
         gives with $q^2_i = 0$
         \bea
         2V_0 &+& s V_4 - \nu V_5 = 0 \co
         \nn
         && sV_1 + \nu V_3 = 0 \co
         \nn
         && sV_7 - \nu V_9 = 0 \per
         \eea
         The second Ward identity $q^\nu_2 V_{\mu \nu} = 0$ is then
         automatically satisfied by Bose symmetry. We are left with
         four independent form factors which we take to be
         \bea
         A = - V_4 \scs
         B = V_9/2s \scs
         C = V_2 \scs
         D = V_3/s \per
         \label{ad2}
 \eea
Insertion into (\ref{ad1}) gives the decomposition Eq. (\ref{na1}) in the text.

\setcounter{equation}{0}
\addtocounter{zahler}{1}
\section{$\gamma \gamma \rightarrow \pi \pi$ to one loop\label{ao}}

         For convenience, we collect here the one-loop expressions for
         the amplitudes $\gamma \gamma \rightarrow \pi^0 \pi^0, \pi^+
         \pi^-$ \cite{bico,dhlin}.

         \subsection{ $\gamma \gamma \rightarrow \pi^+ \pi^-$}

         The matrix element for
         \be
         \gamma(q_1) \gamma(q_2) \rightarrow \pi^+(p_1) \pi^-(p_2)
         \ee
         is given by
         \be
         < \pi^+(p_1) \pi^-(p_2) \mbox{out} \mid \gamma(q_1)
         \gamma(q_2) \mbox{in} > =  i (2 \pi)^4 \delta^4 (P_f - P_i)
         T^C,
         \ee
         with
         \bea
         T^C &=& e^2 \epsilon_1^\mu \epsilon_2^\nu V_{\mu \nu}^C
         \co
         \nn
         V^C_{\mu\nu} &=& i \int dx e^{-i(q_1x + q_2y)}
         < \pi^+ \pi^- \mbox{out} \mid Tj_\mu(x) j_\nu(y) \mid 0 > \nn
         &=&
             A^C T_{1 \mu \nu} +B^C T_{2 \mu \nu}+
         C^CT_{3\mu \nu} + D^CT_{4\mu \nu} \; .
         \eea
        The tensors $T_{i \mu \nu}$ are defined in
         (\ref{na1}).
The helicity amplitudes are
\bea
H_{++}^C &=& A^C + 2(4M_\pi^2 -s) B^C \co \nn
H_{+-}^C &=& \frac{8(M_\pi^4 -tu)}{s}B^C \; .
\label{ao0}
\eea
 The low-energy expansion of the amplitudes $A^C,B^C$
reads \cite{bico} in $SU(2)_L\times SU(2)_R\times U(1)$
with the Condon-Shortley phase convention
          \bea
         A^C&=&-\left\{ \frac{1}{M_\pi^2-t} + \frac{1}{M_\pi^2- u}\right\}
 - \frac{2}{F^2}
         \left \{\bar{G} (s) + \frac{\bar{l}_\Delta}{48\pi^2}
 \right\} + O(E^2) \co
         \nn
         B^C&=& -\frac{1}{2s} \left\{\frac{1}{M_\pi^2-t} + \frac{1}{M_\pi^2-
         u} \right\}+ O(1) \; .
         \eea
The loop-function $\bar{G}(s)$ is discussed in appendix \ref{ai}. We do not
split the result into $A_2,A_4$ etc., because the propagators contain the
physical pion mass at one-loop order--this would make the splitting rather
useless.

         \subsection{$\gamma \gamma \rightarrow \pi^0 \pi^0$}

         The leading term is generated by one-loop diagrams
alone--there is no contribution from ${\cal L}_4$. The result
 is \cite{bico,dhlin}
         \bea
         A_4 &=& \frac{4(s-M^2)}{sF^2} \bar{G}(s)  \co
         \nn
         B_4 &=&0 \; .
         \label{ao1}
         \eea
         CHPT thus predicts the cross section at this order in the energy
         expansion in terms of the two parameters $F$ and $M^2$. The
         amplitude is purely $S$-wave.

         In order to compare the prediction (\ref{ao1}) with the data,
         we identify $F(M)$ with the physical pion decay constant $F_\pi$
         (physical pion mass $M_\pi$),
         as this induces only changes of higher order. The result is
         shown in Fig. 5, where we display the cross
         section $\sigma (s; \mid\cos \theta\mid \leq 0.8)$
         according to Eq. (\ref{ao1}) with a dashed line, together
         with the Crystal Ball data \cite{cball} as a function of
         the center-of-mass energy $E = \sqrt{s}$.
         The cross section is  below the data for $E<
         400$ MeV where the low-energy expansion can be trusted most.
         It also differs by a similar amount from dispersion theoretic
         calculations \cite{goble}-\cite{kaloshin}. An example
         (Fig. 23 in \cite{pehan}) is shown as  dash-dotted lines in the
         figure. The solid line is the two-loop result.

         The amplitude (\ref{ao1}) has the peculiar property that its
         dispersive representation needs a subtraction, although the
         absorptive part vanishes at high energy  sufficiently fast to
         generate a convergent unsubtracted dispersion integral,
         \be
         \mbox{Im} A = O \left(\frac{\ln s}{s}\right), s \rightarrow
         \infty \per
         \ee

Finally, we note that the leading term (\ref{ao1}) approaches a constant in the
chiral limit,
\be
A(s)  =-\frac{1}{4\pi^2F^2} +
O(E^2)\co \hat{m} \rightarrow 0 \co s \neq 0\; .
\ee

\setcounter{equation}{0}
\addtocounter{zahler}{1}
\section{Loop-integrals
\label{ai}}

1. The loop-integral $\bar{G}(s)$ is
\be
         \bar{G}(s) = - \frac{1}{16 \pi^2} \left \{ 1 + \frac{2M^2}{s}
         \int^{1}_{0} \frac{dx}{x} \ln (1-\frac{s}{M^2}x(1-x)) \right \} \; .
         \ee
$\bar{G}$
 is analytic in the complex $s$ - plane, cut along the positive real
axis  for Re $ s \geq 4M^2$. At small $s$,
\be
\bar{G}(s)= \frac{1}{16\pi^2}\sum_{n=1}^\infty {\left(\frac{s}{M^2}\right)}^n
\frac{(n!)^2}
{(n+1)(2n+1)!}\;.
\ee
         The absorptive part is
         \bea
         \mbox{Im} \bar{G} (s) &=& \frac{M^2}{8s \pi }\ln\left\{
\frac{1+\sigma}{1-          \sigma}\right\} \;\;\co  s>4M^2 \co \nn
        \sigma &=& \sqrt{1-4M^2/s}\; .
         \eea
         Use of
         \bea
         \mbox{Li}_2(y) + \mbox{Li}_2 \left( \frac{-y}{1-y} \right) &=& -
         \frac{1}{2} \ln^2 (1-y) \co \nn
         \mbox{Li}_2 (y) &=& - \int^{y}_{0} \frac{dx}{x} \ln (1-x) \co
         \eea
         gives
         \be
          -  16 \pi^2 \bar{G}(s)  =  \left\{
         \begin{array}{lll} 1 & + \frac{M^2}{s} \left(\ln
         \frac{1-\sigma}{1+\sigma} + i \pi \right)^2 &; \qquad 4M^2  \leq s
         \\
         1 & - \frac{4M^2}{s} \mbox{arctg}^2 (\frac{s}{4M^2-s})^\frac{1}{2}
 &;\qquad
0          \leq s \leq 4M^2
         \\
         1 & + \frac{M^2}{s} \ln^2 \frac{\sigma - 1}{\sigma+1} &; \qquad s
\leq
0.          \end{array} \right.
         \ee

         In the text we also need
         \bea
         \stackrel{=}{G} (s) &=& \bar{G} (s) - s\bar{G}'(0) \per
         \eea
2. The loop-integral $\bar{J}(s)$ is
\be
\bar{J}(s) = -\frac{1}{16\pi^2}\int_0^1 dx \ln(1-\frac{s}{M^2}x(1-x)) \; .
\ee
$\bar{J}$ is analytic in the complex $s$ - plane, cut along the positive real
axis  for Re $ s \geq 4M^2$. At small $s$,
\be
\bar{J}(s)= \frac{1}{16\pi^2}\sum_{n=1}^\infty {\left (\frac{s}{M^2}\right)}^n
\frac{(n!)^2}
{n(2n+1)!}\;.
\ee
The absorptive part is
\be
\mbox{Im} \bar{J}(s) = \frac{\sigma}{16\pi}\co \;\; s>4M^2\; .
\ee
Explicitly,
\bea
          16 \pi^2 \bar{J}(s)  =  \left\{
         \begin{array}{lll} &
          \sigma \left(\ln \frac{1-\sigma}{1+\sigma}+i \pi \right) +2 &;
 \qquad 4M^2  \leq s
         \\
          &2-2(\frac{4M^2-s}{s})^\frac{1}{2} \mbox{arctg}
(\frac{s}{4M^2-s})^\frac{1}{2}
 &;\qquad
0          \leq s \leq 4M^2
         \\
          & \sigma \ln \frac{\sigma - 1}{\sigma+1}+2 &; \qquad s  \leq
0.          \end{array} \right.
\eea
In the text we also need
         \bea
         \stackrel{=}{J} (s) &=& \bar{J} (s) - s\bar{J}'(0) \per
         \eea
3. The loop-function $\bar{H}$ is defined in terms of $\bar{G}$ and
$\bar{J}$, \be
\bar{H}(s)=(s-10M^2)\bar{J}(s) +6M^2 \bar{G}(s) \per
\ee

\def\sd{\strut\displaystyle}

\renewcommand{\bea}{\begin{equation} \begin{array}{l}}
\renewcommand{\eea}{\end{array}\end{equation}}

\setcounter{equation}{0}
\addtocounter{zahler}{1}
\section{Low-energy constants from resonance saturation\label{ar}}

Here we give details of the calculation needed to
estimate the renormalized couplings $a_1^r,a_2^r$ and $b^r$.
We consider the exchange of scalar, (axial-)vector and tensor
mesons with mass $M_R \leq 1.2~\GeV$ and follow the procedure
outlined in Ref. \cite{egpr}.
The contributions of the
vector and tensor mesons are evaluated in the framework
of $SU(2)_L \times SU(2)_R \times U(1)$. In order to overcome
the limitations about the
experimental information presently available on the $1^{+-}$ and
$0^{++}$ multiplets we will work in $SU(3)_L \times SU(3)_R$ at large $N_C$.

\subsection{Vector and tensor mesons ($J^{PC}=1^{--},2^{++}$)}

\subsubsection{The lagrangian}
We set
\bea
V_{\mu}(1^{--}) \doteq
\left\{ \begin{array}{lclc}
& \frac{1}{\sqrt 2} & V_{\mu}^i \tau^i \quad ,\quad V=\rho \\
& \frac{1}{\sqrt 2} & V_{\mu}^0 \cdot \unith \quad ,\quad V=\omega ,\phi
\end{array}
\right.
\eea
and have for the kinetic part
\bea \label{are}
{\cal L}_\kin(V,T) =\sd -\frac{1}{4}\sum_{V}
\langle V_{\mu \nu}V^{\mu \nu}-2M_V^2
V_{\mu} V^{\mu}\rangle \\
{}~\\
\sd -\frac{1}{2}
T_{\mu \nu}D^{\mu \nu ;\rho\sigma}T_{\rho\sigma}
\\
\eea
where
\bea
V_{\mu \nu} =\sd D_{\mu}V_{\nu}-D_{\nu}V_{\mu}
\quad ,\\
{}~\\
\sd D_{\mu}V_{\nu}=\partial_{\mu}V_{\nu}+[\Gamma_{\mu}
,V_{\nu} ] \quad , \\
{}~\\
\sd \Gamma_{\mu}=\frac{1}{2} \left\{ u^{\dagger}
\left[
\partial_{\mu}-ir_{\mu} \right] u +
u\left[ \partial_{\mu} -il_{\mu} \right]
u^{\dagger} \right\}
\quad . \\
\label{are00}
\eea
Furthermore, $T^{\mu\nu}=T^{\nu\mu}$ denotes the spin-2 field
for $f_2 (1270)$ with $J^{PC}=2^{++}$, and
\bea
D^{\mu \nu ;\rho\sigma} =\sd (\Box + M_T^2)
\left\{\frac{1}{2} (g^{\mu \rho}
g^{\nu\sigma}+g^{\nu\rho}g^{\mu\sigma}) - g^{\mu \nu}g^{\rho \sigma} \right\}
 \\ ~\\
+\sd g^{\rho \sigma}\partial^{\mu}\partial^{\nu}+
g^{\mu \nu}\partial^{\rho}\partial^{\sigma}
-\frac{1}{2} (g^{\nu\sigma}\partial^{\mu}\partial^{\rho}
+g^{\nu\rho}\partial^{\mu}\partial^{\sigma}
+g^{\mu\sigma} \partial^{\nu}\partial^{\rho}
+g^{\mu\rho}\partial^{\nu}\partial^{\sigma} )
\quad . \\
\eea

The propagator for $T^{\mu\nu}$ is obtained
in the standard manner by exposing the system
 to an external perturbation,
\bea
{\cal L} =\sd {\cal L}_\kin(T)+j^{\mu \nu}T_{\mu \nu}
\quad . \\
\eea
We find
\bea
\sd G^{\mu \nu ;\rho\sigma}(x)
=(2\pi )^{-4}\int \frac{d^4p~ e^{-ipx}}{M_T^2 -p^2-i\epsilon}
P^{\mu \nu ;\rho\sigma}
\quad , \\
{}~\\
\sd P_{\mu\nu ;\rho\sigma} =
\frac{1}{2}(P_{\mu\rho}P_{\nu\sigma}+
P_{\mu\sigma}P_{\nu\rho})-\frac{1}{3}P_{\mu\nu}
P_{\rho\sigma}\quad , \\
{}~\\
\sd P_{\mu\nu}=-g_{\mu\nu}+\frac{p_{\mu}p_{\nu}}{M_T^2} \quad , \\
\eea
with
\bea
\sd D^{\mu \nu ;\rho\sigma}{G_{\rho\sigma ;}}^{\alpha\beta}(x)
=\frac{1}{2}(g^{\mu \alpha}
g^{\nu\beta}+g^{\nu\alpha}g^{\mu\beta} )\delta^4 (x)\; .
\eea

Now consider the couplings of $V$, $T$ to
pions and to photons, linear in the resonance fields.
Since we are interested in terms of order $E^6$ in the effective action, it
suffices to construct interactions which are at most
of order $E^3$. We set
$f_+^{\mu\nu}=2eQF^{\mu\nu}$ where $F^{\mu\nu}$
is the photon field, and take
\bea \label{are0}
\sd {\cal L}_\hint(V,T)=e\epsilon_{\mu \nu\rho\sigma}
F^{\mu \nu}\sum_{V}\left\{
{C}_V^{1} \langle V^{\rho}\{ u^{\sigma},Q\}\rangle
+{C}_V^{2} \langle V^{\rho}u^{\sigma}
\rangle\langle Q\rangle
\right\}\\
{}~\\
\sd +T_{\mu\nu}\sd \{ C_T^{\pi}\Theta_{\pi}^{\mu\nu}
+e^2C_T^{\gamma}\Theta_{\gamma}^{\mu\nu}\}
\quad ,\\
\eea
where $\Theta_{\pi}^{\mu\nu}$
($\Theta_{\gamma}^{\mu\nu}$) is the energy-momentum
tensor of the pion (photon) field,
\bea \label{are1}
\sd \Theta_{\pi}^{\mu\nu}=\frac{F^2}{2}
\langle u^{\mu}u^{\nu}\rangle -\frac{F^2}{4}
g^{\mu\nu}\{\langle u^{\sigma}u_{\sigma}\rangle
+\langle\chi_+ \rangle \}
\quad ,\\
{}~\\
\sd \Theta_{\gamma}^{\mu\nu}=F^{\mu} \,\!\!_{\alpha}F^{\alpha\nu}
+\frac{1}{4}g^{\mu\nu}F^{\rho\sigma}F_{\rho\sigma}
\quad . \\
\label{are01}
\eea

The coupling $V\rightarrow\pi\gamma$ (
$T\rightarrow\pi\pi$, $\gamma\gamma$) has been
considered in ref. \cite{epr} (\cite{wzbijn}) for the case of
nonet fields and $\langle Q\rangle =0$. The interaction
(\ref{are0}) for the spin-2 field differs from the one
proposed in \cite{wzbijn}.
In particular, our amplitude for \gag is smooth at large momenta and
 purely $D$-wave also off the $f_2$-resonance, see below.

\subsubsection{The amplitudes}

 We find for the amplitudes from vector
exchange
\bea
\displaystyle
A_V =C_V\Biggr[
\frac{s-4(t+M_{\pi}^2 )}{M_V^2-t}+
\frac{s-4(u+M_{\pi}^2 )}{M_V^2-u} \Biggr]
\quad ,\\
{}~\\
\displaystyle
B_V =\frac{C_V}{2}\Biggr[
\frac{1}{M_V^2-t}+
\frac{1}{M_V^2-u} \Biggr] \quad
\sem V=\rho ,\omega ,\phi , \\
\eea
where \cite{kaloshin86,ko1,bijdv,bhn,ba2loop}
\begin{equation}\label{GV}
C_V=\frac{3}{\alpha} M_V^3
\frac{\Gamma (V\rightarrow\pi^0\gamma )}{(M_V^2-M_{\pi}^2)^3}
\quad .
\end{equation}
{}From the published \cite{pdg} widths
we obtain
\begin{equation}
C_{\omega} = 0.67\;\GeV^{-2}\scs C_{\rho} = 0.12\;\GeV^{-2}\scs
 C_\phi = 0.2\cdot 10^{-2} \;\GeV^{-2} \; .
\end{equation}
The calculation of the $\rho,\omega$ correction to the $\gamma \gamma
\rightarrow \pi^0 \pi^0$ scattering amplitude has been extended recently to
include both photons off shell \cite{n1}.

Tensor exchange gives\footnote{
The interaction (\ref{are0}) generates tadpole diagrams
where $T^{\rho}_{\rho}$ disappears in the vacuum. These
graphs give rise to additional contributions to the
amplitudes, which however are of higher order in the
energy expansion. We omit them here.}
\bea
\displaystyle
A_T+2(4M_{\pi}^2 -s)B_T=0
\quad ,\\
{}~\\
\displaystyle
B_T=\frac{1}{4}\frac{C_T^{\gamma}C_T^{\pi}}{M_T^2-s} \co
\\
\eea
with
\bea
\displaystyle
\Gamma (T\rightarrow\gamma\gamma )=
\frac{(e^2C_T^{\gamma})^2M_T^3}{80\pi}
\quad ,\\
{}~\\
\displaystyle
\Gamma (T\rightarrow\pi^0\pi^0 )=
\frac{(C_T^{\pi})^2M_T^3}{960\pi}
\left( 1-\frac{4M_{\pi}^2}{M_T^2} \right)^{\frac{5}{2}}
\quad .\\
\eea
{}From the measured widths
$\Gamma (f_2\rightarrow\gamma\gamma )$,
$\Gamma (f_2\rightarrow\pi^0 \pi^0 )=\frac{1}{3}
\Gamma (f_2\rightarrow\pi\pi )$ \cite{pdg}, we find
\bea
 \sd |C_T^{\gamma}|=0.19~\GeV^{-1}
\quad ,\\
{}~\\
\sd |C_T^{\pi}|=9.2~\GeV^{-1}
\quad . \\
\eea

\subsection{Axial-vector and scalar mesons ($J^{PC}=1^{+-},0^{++}$)}
\subsubsection{The lagrangian}

In this paragraph, ${\bar A}$ denotes a $3\times 3$ matrix.
In particular,
\bea
{\bar u}_{\mu}=\sd
-\frac{\partial_{\mu}{\bar{\phi}}}{F}+...
\quad ,\\
{}~\\
\sd {\bar{\phi}}=\sqrt{2}
\left( \begin{array}{ccc}
\frac{\pi^0}{\sqrt{2}}+\frac{\eta}{\sqrt{6}} & \pi^+ & K^+ \\
\pi^- & -\frac{\pi^0}{\sqrt{2}}+\frac{\eta}{\sqrt{6}} & K^0 \\
K^- & {\bar K}^0 & -\frac{2}{\sqrt{6}}\eta
\end{array}\right) \quad ,
\eea
and
${\bar Q}=\frac{1}{3}{\mbox{diag}}(2,-1,-1)$. (We do not include
$\eta$-$\eta^{\prime}$ mixing, as this is of higher order
in the quark mass expansion.) Furthermore, ${\bar B}_\mu $
stands for the axial-vector nonet,
\bea
\displaystyle
{\bar B}_\mu(1^{+-}) =\frac{1}{\sqrt 2}B_{\mu}^i \lambda^i
+\frac{1}{\sqrt 3}B_{\mu}^9 \cdot \unith \quad ,
\eea
and similarly for the scalar nonet ${\bar S}(0^{++})$. The kinetic terms
are
\bea
{\cal L}_\kin({\bar S},\bar B) =\sd
\frac{1}{2}\langle
D^{\mu}{\bar S}D_{\mu }{\bar S}-M_S^2 {\bar S}^2\rangle
{}~\\
\sd -\frac{1}{4}
\langle{\bar B}_{\mu \nu}{\bar B}^{\mu \nu}-2M_B^2
{\bar B}_{\mu} {\bar B}^{\mu}\rangle
\quad , \\
\eea
where $M_B$ denotes the common nonet mass, and the
covariant derivatives are the $SU(3)$ version of
(\ref{are00}). The couplings to pions and to photons are \cite{egpr,wzbijn,ko2}
\bea
\sd {\cal L}_\hint=e^2
C_S^{\gamma}F_{\mu\nu}F^{\mu\nu}\langle
{\bar Q}^2 {\bar S}\rangle
+C_S^d \langle {\bar S}{\bar u}^{\mu}{\bar u}_{\mu}
\rangle
\\
{}~\\
\sd +C_S^m \langle {\bar S}{\bar \chi}_+ \rangle
+eC_B F_{\mu\nu} \langle {\bar B}^{\mu}
\{ {\bar Q},{\bar u}^{\nu}\} \rangle
\quad . \\
\label{av1}
\eea
\subsubsection{The amplitudes}

We find\footnote{
The interaction (\ref{av1}) generates tadpole diagrams
where $\langle {\bar S}\rangle$ disappears in the vacuum. These
graphs give rise to additional contributions to the
amplitudes, which however are of higher order in the
energy expansion. We omit them here.} for the contribution from
the scalar exchange \cite{wzbijn}
\bea
\displaystyle
A_S=\frac{20C_S^{\gamma}}{9F_{\pi}^2(M_S^2-s)}
[sC_S^d+2M_{\pi}^2(C_S^m-C_S^d)] \quad
,\\
{}~\\
\displaystyle
B_S=0\quad , \\
\eea
with
\bea
\displaystyle
\Gamma (a_0\rightarrow\gamma\gamma )=
\frac{(e^2C_S^{\gamma})^2M_S^3}{72\pi}
\quad .\\
\eea
The couplings  $C_S^d$ and $C_S^m$ have been determined
in \cite{egpr},
\bea
\sd |C_S^d|=3.2\cdot 10^{-2}~\GeV\quad ,
\\
{}~\\
\sd |C_S^m|=4.2\cdot 10^{-2}~\GeV
\quad ,
\\
{}~\\
\sd C_S^d C_S^m >0\quad .\\
\eea
The quantity $|C_S^{\gamma}|$ is difficult to
estimate. Here we content ourselves with a rough estimate
by relating it to the decay $a_0\rightarrow\gamma\gamma$
via
\bea
\displaystyle
\Gamma (a_0\rightarrow\gamma\gamma)=r
\frac{\Gamma_\tot}{\Gamma (a_0\rightarrow\pi^0\eta )}
\quad ,
\eea
where $r=0.24~\keV$ \cite{pdg} with a sizeable error. We
furthermore assume that the decay
$a_0\rightarrow\pi^0\eta$ accounts for
all of $\Gamma_\tot$
\bea
\Gamma_\tot=
\Gamma (a_0\rightarrow\pi^0\eta )
\quad .
\eea
Hence we obtain
\bea
 |C_S^{\gamma}|=8.2 \cdot 10^{-2}~\GeV^{-1}
\quad .
\eea

For the contribution from the axial nonet exchange \cite{kaloshin86,ko2}
 we find

\bea
\displaystyle
A_B =C_B\Biggr[
\frac{s+4(t-M_{\pi}^2 )}{M_B^2-t}+
\frac{s+4(u-M_{\pi}^2 )}{M_B^2-u} \Biggr]
\quad ,\\
{}~\\
\displaystyle
B_B =\frac{C_B}{2}\Biggr[
\frac{1}{M_B^2-t}+
\frac{1}{M_B^2-u} \Biggr]
 \\
\eea
with
\begin{equation}\label{GB}
C_B=\frac{30}{\alpha} M_B^3
\frac{\Gamma (b_1\rightarrow\pi^+\gamma )}{(M_B^2-M_{\pi}^2)^3}
\end{equation}
or
\be
C_B= 0.53~\GeV^{-2} \quad .
\ee

\subsection{Expressions at low energies}

At low energies, the above contributions from $V,B,S$ and $T$ sum up to
\bea\label{final}
\displaystyle
A_6=\bar{a}_1M_{\pi}^2 +\bar{a}_2s \quad
,\\
{}~\\
\displaystyle
B_6=\bar{b} \quad ,\\
{}~\\
\displaystyle
\bar{a}_1=-16\sum_{V=\rho ,\omega ,\phi}\frac{C_V}{M_V^2}
\pm \left( \frac{40}{9}
\frac{|C_S^{\gamma}(C_S^m-C_S^d)|}{F_{\pi}^2M_S^2}
-2\frac{|C_T^{\gamma}C_T^{\pi}|}{M_T^2}
\right)
\quad ,\\
{}~\\
\displaystyle
\bar{a}_2=6\sum_{V=\rho ,\omega ,\phi}\frac{C_V}{M_V^2}
-2\frac{C_B}{M_B^2}\pm\left( \frac{20}{9}
\frac{|C_S^{\gamma}C_S^d|}{F_{\pi}^2M_S^2}
+\frac{1}{2}\frac{|C_T^{\gamma}C_T^{\pi}|}{M_T^2}
\right) \quad ,\\
{}~\\
\displaystyle
\bar{b}=\sum_{V=\rho ,\omega ,\phi}\frac{C_V}{M_V^2}
+\frac{C_B}{M_B^2}\pm\frac{1}{4}
\frac{|C_T^{\gamma}C_T^{\pi}|}{M_T^2}
\quad .\\
\eea
In table 2 we list the contributions from the individual resonances to the
dimensionless parameters
$(16\pi^2F_\pi^2)^2(\bar{a}_1; \bar{a}_2; \bar{b})$.
We used the following values for the resonance masses:
$M_{\omega}=782~\MeV$,
$M_{\rho}=768~\MeV$,
$M_{\phi}=1020~\MeV$,
$M_{B}=1232~\MeV$,
$M_{S}=983~\MeV$,
$M_{T}=1275~\MeV$.

\newpage

{\Large {\bf{Figure captions}}}

\vspace{1cm}

Fig. 1. Mandelstam plane showing the three related physical
regions. $s$-channel: \ggpp, $t$-and $u$-channel: \gpgp.
 We indicate the threshold for
\ggpp(\gpgp) by A (P).
The shaded lines at $s,t,u$ = $4M_\pi^2$ indicate the presence
of a branch-point in the amplitude, generated by two-pion intermediate states.

Fig. 2. The graphs at order $E^6$ in the chiral expansion. These graphs
correspond to the  terms in Eq. (\ref{le11}). The framed symbols $I$ stand
 for the
vertices in ${\cal{L}}_I$. We indicate with the solid-dashed lines the
propagator
$D_2^{-1}$.

Fig. 3. One class of Feynman diagrams which contribute to \ggppz
. The dashed
(solid) lines stand for neutral (charged) pions. The four-point function on the
right-hand side is the elastic $\pi\pi$ scattering amplitude at one-loop
accuracy in $d$-dimensions (with two legs off-shell),
 and the symbol $d^Dl$ stands for integration over
internal lines with weight (\ref{fa0}).

Fig. 4. Graphs which contribute to $\Delta_{A,B}$ in the unitary part
$U_{A,B}$ in Eqs. (\ref{ac2},\ref{ac3}). In Fig. 4a, the dash-dotted lines
surround
the diagram which we represent in the dispersive manner (\ref{fa1a}). The graph
Fig. 4b is called "acnode graph" \cite{elop}.

Fig. 5. The \ggppz $\;$cross section $\sigma(|\cos\theta|\leq Z)$ as a function
of the center-of-mass energy $E$ at $Z=0.8$, together with the data
 from the Crystal Ball
experiment \cite{cball}. The solid line is the full two-loop result,
and the dashed line results from the one-loop calculation \cite{bico,dhlin}.
 The band denoted by
the dash-dotted lines is the result of the dispersive calculation by
Pennington (Fig. 23 in \cite{pehan}).

Fig. 6. The dependence of the \ggppz $\;$ cross section
 $\sigma(|\cos\theta|\leq Z)$
on the constants $\bar{l}_i$, at $Z=0.8$. The solid line denotes the two-loop
result with
the standard values for $\bar{l}_i$ displayed in table 1 (without
contributions from resonance exchange, and with $\Delta_{A,B}=0$), whereas  the
dashed
line is evaluated at  $\bar{l}_i=0$. The dash-dotted
line has $\bar{l}_{1,3}=0$ and the other $\bar{l}_i$ at their standard values.

Fig. 7. Real and imaginary part of the amplitudes $\pm 10^2M_\pi^2H_{+\pm}$
at $t=u$. The
solid line is for $10^2M_\pi^2H_{++}$. It incorporates all contributions
except $\Delta_{A,B}$. The dashed line is the same amplitude for the one-loop
case, and the dash-dotted line is for  $-10^2M_\pi^2H_{+-}$ with
 the same input as
the solid line. Finally, the crosses refer to the center-of-mass energy of
the $\pi^0 \pi^0$ system in $100$ MeV steps.

Fig. 8. The \ggppz $\;$ amplitude as a function of $s/M_\pi^2$ at $t=u$.
 For $s \leq
4M_\pi^2$ the quantity shown is $10^2M_\pi^2H_{++}$ and for $s\geq 4M_\pi^2$
we display  $10^2M_\pi^2|H_{++}|$. The solid line is the result of the
two-loop calculation and the dashed line is the one-loop result. The symbols
$(\diamond,+)$ refer to the work of Pennington \cite{pehan}: $\diamond (+)$ is
from Fig. 19 (23) in that article.

Fig. 9. The uncertainty in the value of the $\gamma \gamma \rightarrow
\pi^0 \pi^0$ cross section
 $\sigma(|\cos\theta|\leq Z)$
 at $Z=0.8$, evaluated from the two-loop amplitude at $\Delta_{A,B}=0$. The
data
are from the Crystal Ball
experiment \cite{cball}.
The dashed lines  embrace the region generated by
assigning all possible combinations of signs to the systematic errors
in the couplings $h_\pm^r$ and $h_s^r$ according to Eq. (\ref{le20}).
 The dash-dotted
line corresponds to the central values in (\ref{le20}).
 Above $E \simeq 400$ MeV,
 the major part of the uncertainty in the cross section is
generated by the error in $h_s^r$.

Fig. 10. The Compton cross section \gpgpz $\;$ as a function of the
center-of-mass
energy $E_{\gamma \pi}$. The solid line is the result of the two-loop
calculation and the dashed line is the one-loop result. The dash-dotted
line refers to the two-loop calculation at $h_\pm^r=h_s^r=0$, and
the dotted line has $H_{+-}=0$ in a full two-loop calculation.

Fig. 11. The \ggppz $\;$ cross section $\sigma(|\cos\theta|\leq Z)$ as a
function of
the center-of-mass energy at $Z=0.8$, with the data from the Crystal
 Ball \cite{cball}
experiment. The solid line is the two-loop result, whereas the dashed line
is taken from the dispersive analysis of Donoghue and Holstein
 (Fig. 2 in \cite{dohod}).
\end{document}